\documentclass[journal=jacsat,manuscript=article]{achemso}

\usepackage[version=3]{mhchem} 
\usepackage{xcolor}
\usepackage{comment}
\usepackage{xr}
\usepackage{soul}

\SectionNumbersOn
\title[An \textsf{achemso} demo]
  {Dissipative self-assembly of patchy particles under nonequilibrium drive: a computational study}

\author{Shubhadeep Nag}
\affiliation{Department of Biomedical Engineering, Faculty of Engineering, Tel Aviv University, Tel Aviv 69978, Israel}
\author{Gili Bisker}
\affiliation {Department of Biomedical Engineering, Faculty of Engineering, Tel Aviv University, Tel Aviv 69978, Israel}
\alsoaffiliation[Second University]
{The Center for Physics and Chemistry of Living Systems, Tel Aviv University, Tel Aviv 6997801, Israel}
\alsoaffiliation[Third University]{The Center for Nanoscience and Nanotechnology, Tel Aviv University, Tel Aviv 6997801, Israel}
\alsoaffiliation[Fourth University]{The Center for Light-Matter Interaction, Tel Aviv University, Tel Aviv 6997801, Israel}
\alsoaffiliation[Fifth University]{The Center for Computational Molecular and Materials Science, Tel Aviv University, Tel Aviv 6997801, Israel}
\email{bisker@tauex.tau.ac.il}

\abbreviations{IR,NMR,UV}

\begin{document}


\begin{abstract}

Inspired by biology and implemented using nanotechnology, the self-assembly of patchy particles has emerged as a pivotal mechanism for constructing complex structures that mimic natural systems with diverse functionalities. Here, we explore the dissipative self-assembly of patchy particles under nonequilibrium conditions, with the aim of overcoming the constraints imposed by equilibrium assembly. Utilizing extensive Monte Carlo (MC) and Molecular Dynamics (MD) simulations, we provide insight into the effects of external forces that mirror natural and chemical processes on the assembly rates and the stability of the resulting assemblies comprising $8$, $10$, and $13$ patchy particles. Implemented by a favorable bond-promoting drive in MC or a pulsed square wave potential in MD, our simulations reveal the role these external drives play in accelerating assembly kinetics and enhancing structural stability, evidenced by a decrease in the time to first assembly and an increase in the duration the system remains in an assembled state. Through the analysis of an order parameter, entropy production, bond dynamics, and interparticle forces, we unravel the underlying mechanisms driving these advancements. 
We also validated our key findings by simulating a larger system of $100$ patchy particles.
Our comprehensive results not only shed light on the impact of external stimuli on self-assembly processes but also open a promising pathway for expanding the application by leveraging patchy particles for novel nanostructures.

\end{abstract}


\section{Introduction}

Self-assembly is a fundamental process in nature, in which particles within a bounded system spontaneously organize into an ordered structure, directed by intermolecular interactions. 
This phenomenon is not only pivotal in a myriad of biological processes, such as protein folding and cell membrane formation, but it also plays a crucial role in advancing nanotechnology, facilitating the creation of sophisticated structures across multiple scales \cite{Whitesides2002a, Reinhardt2016, Truskett2017, Trusket2017b, Reinhardt2018, Liu2021, Bassani2024}. 
Particularly in nanoscience, self-assembly enables the development of innovative applications, ranging from drug delivery systems, exemplified by liposomes formed from lipid bilayers \cite{Gao2013, Verma2013, Filipczak2020}, to the construction of advanced materials for photonic and electronic sensing devices using organized gold nanoparticles \cite{Shaw2011, Yang2017, Bian2018}. 
These diverse applications accentuate the importance of understanding and modeling self-assembly processes.

Self-assembly is categorized into two primary types: static and dissipative. 
In static self-assembly, the formed structure does not rely on any external source of energy to maintain itself, and the system minimizes its free energy under equilibrium conditions \cite{Whitesides2002a}. 
Conversely, dissipative self-assembly occurs under
nonequilibrium conditions, often at the expense of external energy sources, and the system dissipates energy to maintain its structure \cite{Fialkowski2006, England2015, Bisker2020,Makey2020}. 
This nonequilibrium behavior is driven by a continuous energy supply from various sources such as light \cite{Liu2019}, chemical fuels \cite{Cheng2023, Raphael2019}, electric \cite{Wiewiorski2023}, and magnetic fields \cite{Park2022}, resulting in a non-zero difference between the forward and backward fluxes, thereby activating individual components and form the desired structures.

The theory of dissipative states, first introduced by Prigogine \cite{Prigogine1978}, is continuously developing and is found to be relevant to many biological phenomena \cite{England2015, Marshland2015, Marsland2018, Nguyen2021, Pavel2021, England2022}.
Particularly, the thermodynamics of continuous energy-driven structure forming systems in nonequilibrium conditions has been a rapidly developing field with the introduction of various design principles to model dissipative self-assembly  \cite{Penocchio2019, Tociu2019,  Penocchio2021, Korbel2021, Trubiano2021, Gartner2022, Gartner2024}. 
One major goal of these studies is to overcome the bottlenecks or trade-offs imposed by equilibrium or static self-assemblies.

Equilibrium self-assembly raises many unavoidable trade-offs, each reflecting the intricate balance systems must achieve. For example,  a system might navigate a trade-off between kinetic and thermodynamic factors, such that it may become confined in kinetically trapped states, failing to achieve the most thermodynamically favorable configuration. This can result in suboptimal or inefficient assembly processes \cite{Marsland2018}.
Numerous design principles have been suggested to overcome kinetically trapped states \cite{Hagan2011, Michaels2017}. To circumvent these limitations, researchers have explored diverse self-assembly strategies, employing nonreciprocal interactions and driving systems from arrested dynamics towards lower-energy configurations \cite{Murugan2015, Osat2023a}. 
This strategy leverages nonequilibrium dynamics and programmable interactions, opening up new avenues for the creation of dynamic, self-organizing structures across different scales - from nucleic acids to colloidal particles, facilitating automated control over structural transitions in living systems \cite{osat2023b}. Additionally, Boekhoven and colleagues introduced a dissipative self-assembling system that utilizes external fuel, analyzing the associated entropy production and lost work \cite{Boekhoven2010, Koper2013}.
England explored the complex thermodynamics of nonequilibrium processes, crucial in understanding self-assembly and self-replication in biological systems \cite{England2022}. 
His work delves into the challenges of modeling {nonequilibrium} scenarios, especially in driven self-assembly, proposing mechanisms for self-organization via energy dissipation \cite{England2013, England2015, Kedia2023}.

Biological systems also exhibit these two types of assemblies. 
For instance, static self-assembly orchestrates the formation of functional hemoglobin proteins from hemoglobin polypeptides and functional ribosomes from the combination of RNA and ribosomal proteins \cite{Subramani2018}. 
In contrast, dissipative self-assembly plays a crucial role in cell division and macromolecular assembly, such as tubulin dimer organization into microtubules facilitated by Guanosine triphosphate (GTP) hydrolysis, and chaperone-mediated macromolecule formation through Adenosine triphosphate (ATP) hydrolysis \cite{Sartori2020, vale1994, Mayer2005, Goloubinoff2018}. 
Additionally, it is integral to active matter, driving self-organization in cellular cytoskeletons and bacterial colonies using external energy sources \cite{Fletcher2010}.
Furthermore, experimental studies have highlighted the importance of ATP and GTP induced self-assembly systems \cite{Vopel2010, Deng2020}, offering profound insights into the development of synthetic cells, prebiotic systems, and nanosystems  \cite{Ma2021, Dzieciol2012}. 
These advancements enhance our understanding of dissipative self-assembly modeling, crucial for deepening our comprehension of biological structures and offering significant potential in biochemical and materials sciences \cite{Cheng2015, Pezzato2017, Rao2023}.

The aforementioned example of the organization of tubulin dimers into microtubules, facilitated by GTP  hydrolysis, serves as a primary example of dissipative self-assembly found in our cell \cite{vale1994}.
These microtubules feature a cylindrical shape with a hollow core, appearing as a ring of $13$ dimers when viewed perpendicularly to the core's radius \cite{Bechstedt2012}.
Similar ring-like self-assembly structures are found in circular DNA and cyclic protein complexes, which are crucial for understanding complex biological assembly processes \cite{Insua20, Zhao22, Cabezon2023}. The significance of these rings extends into nanoscience, serving as foundational blueprints for developing self-assembled nanorings \cite{Li10, Mews22}. These nanorings, far from being merely minuscule structures, are pivotal building blocks for transformative applications in drug delivery, nanoelectronics, and photonic devices \cite{Dunn23}. Therefore, modeling such ring-like structures, and utilizing the external sources of energy helps understand biological phenomena and design advanced materials.

Computational studies play a crucial role in advancing our understanding of structures created through dissipative self-assembly, offering deeper microscopic insights. Within this framework, patchy particles, characterized by distinct patches that enable selective, anisotropic binding, emerge as a fundamental model system \cite{Li2020}. The theoretical foundation for these interactions traces back to seminal theories proposed by Boltzmann \cite{Boltzmann1905}, and were first practically demonstrated with the synthesis of Janus particles, a notable subclass of patchy particles, in the late $1980$s \cite{Casagrande1989}. The specific term `patchy particle' was later formally coined \cite{Glotzer2004}, and the introduction of this concept has unlocked vast possibilities in predicting crystal structures, thereby significantly contributing to the advancement of nanoscience and nanofabrication. This breakthrough has sparked a wave of research, exploring the broad applications and potential of patchy particles in various domains \cite{Hong2006, Pawar2009, Gong2017, Lotierzo2019, Paul2020, Virk2023}.
For example, Gong \emph{et.~al.} synthesized patchy particles using colloidal fusion \cite{Gong2017}. 
Additional study demonstrated the synthesis of Janus particles 
using evaporative deposition of preferential materials on spheroid, changing 
the properties of one hemispheroid \cite{Shah2013}.
Chang and others showed a simple yet effective technique by using seed-mediated heterogeneous nucleation to synthesize patchy particles \cite{Chang2021}.
These patchy particles have been found useful in many potential applications such as targeted drug delivery \cite{Langer2004, Wong2019}, photonic crystals \cite{Liddell2003, Rao2020, Cai2021, Flavell2023}, phoretic motors \cite{Pawar2010}, and more.
Recent explorations into the self-assembly of patchy particles under external stimuli, such as electric \cite{Song2015} and magnetic fields \cite{Shields2013}, have further highlighted their role in advancing new material designs.

In computational research, patchy particles have been extensively employed to simulate diverse systems, such as protein droplets \cite{Nguemaha2018, Gnan2019} and colloidal assemblies \cite{Li2020}, enabling a profound exploration of phenomena like crystal nucleation and growth \cite{Sciortino2021, Beneduce2023}, liquid-liquid phase transitions \cite{Nguemaha2018}, and protein folding \cite{McMullen2018}, among others \cite{Doye2007, Doye2010, Doye2021}. By tuning the characteristics of patches, models based on patchy particles have offered insights into the phase behavior of colloids \cite{Bianchi2006}.
Specifically, Nguemaha \emph{et.~al.} utilized Monte Carlo (MC) simulations with two-component patchy particles to investigate RNA's role in the liquid-liquid phase separation observed within membrane-less organelles, focusing on protein mixtures that incorporate regulatory components \cite{Nguemaha2018}. Similarly, Gnan, Sciortino, and Zaccarelli's coarse-grain modeling of proteins using patchy particles has shed light on protein phase behavior \cite{Gnan2019}. Espinosa \emph{et.~al.} have also contributed by employing Molecular Dynamics (MD) simulations with patchy particles to demonstrate deviations from the law of rectilinear diameter through an examination of liquid-vapor coexistence phase behavior \cite{Espinosa2019}. 
Thus, the well-established understanding of patchy particles positions them as an ideal model system, which aims to leverage these insights further to understand biological self-assembly phenomena and pioneer new materials.

Designing such new materials requires the mitigation of the aforementioned trade-offs effectively. 
One unique approach to achieve this is by driving the system into non-equilibrium conditions, for example, through the introduction of a self-healing driving force \cite{Bisker2018}.
This proposed model of dissipative self-assembly is based on information from equilibrium self-assembly and biological processes such as the use of chemical fuel to organize structures with specific functions similar to the way microtubules are formed by $\alpha$-tubulin and $\beta$-tubulin through GTPase activity \cite{Zhou2023}. 
Inspired by these processes, it was demonstrated that nonequilibrium driving enhances dimer-based self-assembly, enabling smaller critical seeds and improved stability of target structures compared to equilibrium scenarios \cite{Ben2021}.
Using a similar model, the stochastic landscape method was proposed for improving the predictive power for first assembly times, offering a quantitative framework for understanding and controlling nonequilibrium self-assembly processes \cite{Faran2023}. Recently, it has been used to classify protein states \cite{Faran2024}.

In the present study, we aim to expand this approach of nonequilibrium driving force by undertaking a detailed exploration of self-assembly mechanisms using patchy particles in {three-dimension ($3$D)}, employing a comprehensive simulation approach.
Our investigation begins with equilibrium MC simulations across three distinct systems, each comprising $8$, $10$, and $13$ patchy particles, respectively, where each particle has two possible internal states.
These particles are specifically designed to achieve target structures of $n$-sided polygons or $n$-ring configurations, comprising all the particles in the system. 
We examine the formation dynamics and stability of the self-assembled structures, particularly in relation to the varying interaction energies between patches, and identify a domain of equilibrium trade-offs,
delineating a complex relationship between assembly time and structural stability. 
To overcome these trade-offs, an external bias is introduced, 
shifting the dynamics to the nonequilibrium regime,
inspired by biological systems that exploit energy molecules like ATP and GTP 
thus mimicking the continuous energy supply essential for sustaining assembly processes in 
living systems.\cite{Heinonen2012}.
A rigorous thermodynamic analysis encompassing entropy production and total energy computations shows
how this externally applied drive enables the system to form the target structure faster compared to the equilibrium case.
Complementing the MC results, we perform MD simulations for systems of $8$ and $10$ patchy particles under both equilibrium and nonequilibrium conditions.
For nonequilibrium MD simulations, a square wave potential is applied periodically as an external drive, where we further analyze the bond formation and breakage statistics, the underlying forces, and an order parameter to distill the underlying mechanism governing the dissipative self-assembly.
We demonstrate how the oscillating potential induces momentary forces that accelerate the assembly, allowing the system to rapidly achieve the target structure.
By revealing the nuanced dynamics of dissipative self-assembly with patchy particles, this study highlights the critical role of external energy in guiding assembly processes. Our insights offer a foundation for novel approaches in biophysics and material engineering, setting the stage for advancements in the design and control of self-assembling systems.

\section{Computational Framework}
\subsection{Simulation Cells}
\label{simdetails}

\begin{figure}
    \centering
    \includegraphics[width=14cm]{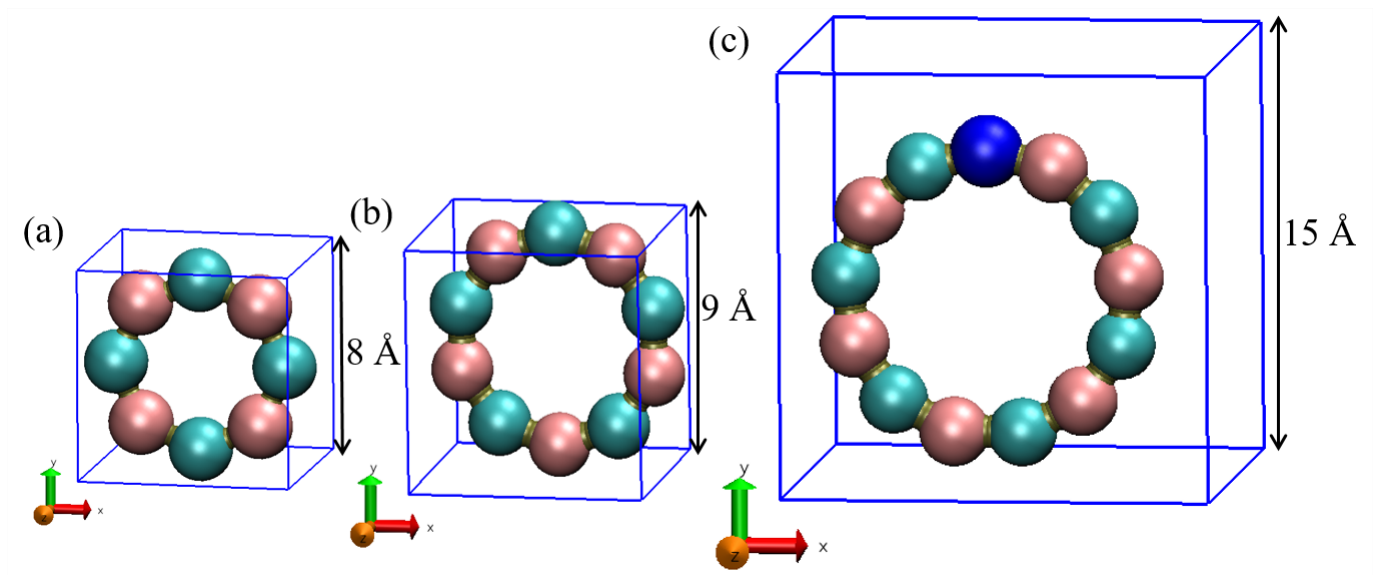}
    \caption{Target structures of (a) $8$-patchy particle system, (b) $10$-patchy particle system, and (c) $13$-patchy particle system. 
    Each target structure is presented within its cubic simulation cell, with lattice parameters set to $8$ \AA, $9$ \AA, and $15$ \AA\ for the $8$, $10$, and $13$-particle systems. In the $8$ and $10$-particle systems, particles in states $\alpha$ and $\beta$ are represented by pink and cyan colors, respectively. For the $13$-patchy particle system, the $\gamma$ state appears in blue.}
    \label{target}
\end{figure}

MC simulations were conducted for three systems, comprising $8$, $10$, and $13$ patchy particles, whereas MD simulations were conducted for systems with $8$ and $10$ patchy particles. Each particle was designed with two surface patches and varied states depending on the system. 
In the systems with $8$ and $10$ patchy particles, we divided the particles into two distinct states, labeled $\alpha$ and $\beta$, with equal distribution among the particle populations. 
Conversely, in the system comprising $13$ patchy particles, we introduced a third state, $\gamma$, alongside the existing $\alpha$ and $\beta$ states. 
In this scenario, six particles were assigned to the $\alpha$ state and another six to the $\beta$ state, with the remaining particle embodying the $\gamma$ state. 
Our objective was to form target structures that emulate polygons with $8$, $10$, and $13$ sides, corresponding to each system (Fig.~\ref{target}). 
To realize these structures, the patches on each particle were meticulously positioned to ensure 
that the angle between two lines - each extending from a patch to the particle's center - matched the interior angles of an $8$-sided, $10$-sided, and $13$-sided polygon, respective to each system.
These angles are $2.356$ rad, $2.513$ rad, and $2.658$ rad for $8$, $10$, and $13$ patchy particle system respectively.
Furthermore, the size of the cubic simulation boxes was carefully chosen based on the number of patchy particles to be $8$ \AA, $9$ \AA, and $15$ \AA\ for the $8$-particle, $10$-particle, and $13$-particle systems, respectively.
In MD simulation, we simulated $8$ and $10$ particle systems with the same simulation parameters as described above, except that the particles had only one internal state, $\alpha$.

\subsection{Interaction Potential of Patchy Particles}

Patchy particle simulations have been a versatile tool for modeling a wide range of structures, ranging from proteins and DNA to small molecules and atoms \cite{Rovigatti2018}.
Kern and Frenkel proposed an approach for simulating patchy particles by modeling the bead of the patchy particles with hard-sphere potential and patches with both square well potential along with an angular dependent potential.\cite{Kern2003}
This model has been shown effective in simulating many cases of patchy particles \cite{Zhang2005, Bolhuis2015, Panagiotopoulos2016,  Bolhuis2021, Sciortino2021}.

\begin{figure}
    \centering
    \includegraphics[width=10cm]{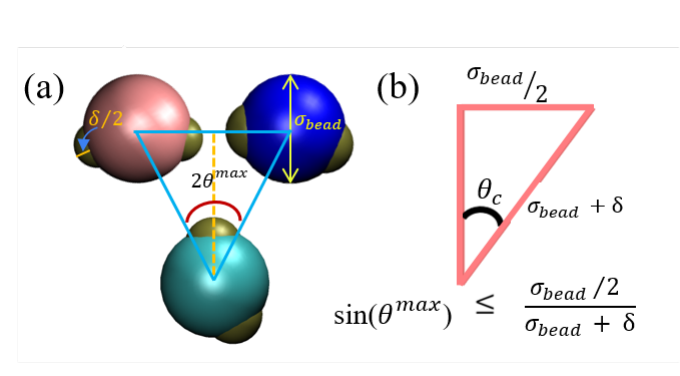}
    \caption
    {Patchy particles model. (a) A schematic where three patchy particles may potentially interact with each other, resulting in an unwanted case of multiple bonds per patch. (b) A specific condition is implemented to prevent more than a single bond per patch (see details in the main text). This figure is a schematic representation created to visualize the concepts described, with dimensions and proportions aligned with those used in our simulations.}
    \label{singlebond}
\end{figure}

We build upon the Kern-Frenkel approach and apply it to our system of $8$, $10$, and $13$ patchy particles.
Based on this approach, each patchy particle is featured with patches of an angular width of $2\theta^{max}$ centered to the core of the bead and length $\delta/2$ (Fig.~\ref{singlebond}(a)).
In this framework, the total interaction energy of a patchy particle, $U_{Particle}$, is:
\begin{equation}
    U_{Particle} = U_{bead} + U_{patch}
    \label{model1}
\end{equation}
where $U_{bead}$ and $U_{patch}$ are the energies associated with the bead and the patch, respectively.

The expression of $U_{bead}$ is modeled as Lennard-Jones (LJ) interaction: 
\begin{equation}
    \label{repulsive}
    U_{bead} = 
    \begin{cases} 
    4\epsilon_{bead} \Bigg[\Big(\frac{\sigma_{bead}}{r_{ij}}\Big)^{12} -\Big(\frac{\sigma_{bead}}{r_{ij}}\Big)^{6} \Bigg] ,  \ \ \ if \ \  
\sigma_{bead} < r_{ij} < r_{c} \\
              0 , \ \ \ \ otherwise\\
    \end{cases}
\end{equation}

\noindent where $\epsilon_{bead}$ is the depth of the potential well of the bead and $\sigma_{bead}$ is its van der Waals (vdW) radius. These two values are set at $0.75$ kJ/mol and $2$ \AA\ throughout the simulation. When the distance between two patchy particles $i$ and $j$, denoted as  $r_{ij}$, is smaller than the cut-off distance, $r_{c}$, the particles interact.
Here, $r_{c}$ is taken as $5$ \AA.

In the case of MC simulation of $8$ and $10$ patchy particles, the two different states $\alpha$ and $\beta$ attract each other, whereas the same states do not interact with each other ($U_{patch}^{\alpha\alpha}=0$,$U_{patch}^{\beta\beta}=0$). 
The patch energy {$U_{patch}^{\alpha\beta}$ between $\alpha$ and $\beta$ states of} the $8$ and $10$ patchy particle systems is:
\begin{equation}
\label{2_state_U}
    U^{\alpha\beta}_{patch} = \epsilon_{patch}\sum_{\alpha, \beta=1}^{2} f(\Vec{r}_{ij}, \hat{n}^{\alpha}_{i}, \hat{n}^{\beta}_{j}) , \ \  if \  \ \ \sigma_{bead} < r_{ij} < \sigma_{bead}+\delta \\
\end{equation}
where $\epsilon_{patch}$ denotes the depth of the potential between patches of the particle, 
$\Vec{r}_{ij}$ is the vector connecting the center of particle $i^{th}$ to the center of particle $j^{th}$, $\hat{n}^{\alpha}_{j}$ or $\hat{n}^{\beta}_{j}$ denote the unit vector connecting the center of the bead and the center of patches of the patchy particle of state $\alpha$ or $\beta$, respectively, and $f(\Vec{r}_{ij}, \hat{n}^{\alpha}_{i}, \hat{n}^{\beta}_{i})$ is the angular dependence in the interaction between patches, and which is given by:

\begin{equation}
\label{2_state_f}
f(\Vec{r}_{ij}, \hat{n}^{\alpha}_{i}, \hat{n}^{\beta}_{j}) = 
    \begin{cases}
        1, \ \  if \  \ \Vec{r}_{ij}\cdot\hat{n}^{\alpha}_{j} > \text{cos} (\theta^{max}) \  \& \  \Vec{r}_{ji}\cdot\hat{n}^{\beta}_{j} > \text{cos} (\theta^{max})\\
        0, \ \  otherwise
    \end{cases}
\end{equation}
This condition ensures that the patchy interaction potential is activated only when the angle between $\vec{r}_{ij}$, and the unit vectors $\hat{n}^{\alpha}_{i}$ and $\hat{n}^{\beta}_{j}$, representing the orientation of their respective patches, falls within half of the angular width of the patch, $\theta^{max}$. 
 Specifically, this occurs when both $\vec{r}_{ij} \cdot \hat{n}^{\alpha}_{i}$ and $\vec{r}_{ij} \cdot \hat{n}^{\beta}_{j}$ are  {larger} 
 than the cosine of $\theta^{max}$. When these conditions are satisfied, meaning the patches are correctly aligned, the angular dependency term equals $1$, and the interaction potential $U_{patch}$ of Eq.~\ref{model1}  contributes to the total interaction energy of the patchy particle.

For the $13$ patchy particles system, there are three internal states: $\alpha$, $\beta$, and $\gamma$. The $\alpha$ and $\beta$ states interact with each other similar to the $8$ and $10$ particle cases. The interaction between the $\gamma$ state and the $\alpha$ or $\beta$ states is the energy $U_{patch}^{\alpha / \beta \gamma}$, with a similar form as Eq.~\ref{2_state_U} and the corresponding angular dependence $f(\Vec{r}{ij}, \hat{n}^{\alpha / \beta}_{i}, \hat{n}^{\gamma}_{j})$ similar to Eq.~\ref{2_state_f}.
This ensures that the interaction potential $U_{patch}$ is activated only when both the angle between $\Vec{r}_{ij}$ and the unit vector $\hat{n}^{\alpha / \beta}_{i}$, and the angle between $\Vec{r}_{ji}$ and $\hat{n}^{\gamma}_{j}$, fall within half of the angular width of the patch, $\theta^{max}$. Here, $\hat{n}^{\gamma}_{j}$ denotes the unit vector connecting the center of the bead and the center of patches of the $\gamma$ patchy particle.

To ensure a single bond per patch \cite{Smallenburg2013, Rovigatti2018}, the length of the patch must be related to the angular width by the inequality $sin(\theta^{max}) \leq (\sigma_{bead}/2)/(\sigma_{bead}+\delta)$. 
To satisfy this condition, we chose $\theta^{max}$ to be $0.35$ radians. 
The schematic of this condition is shown in Fig.~\ref{singlebond}(b).

\subsection{Initial Conditions}
\label{ini_condition}

To rigorously assess the inherent trade-offs associated with equilibrium simulations and to determine how an external bias can mitigate these limitations, our investigation centers on two critical metrics, namely, the time until the first self-assembly event, $T_{fas}$, and time the system remains in the assembled state as a proxy for target stability, $T_{stable}$.
Accordingly, our approach in both MC and MD simulations encompassed two distinct methodologies in which the initial conditions were set to be random, for analyzing $T_{fas}$, or as the target structure, for analyzing $T_{stable}$.
In the former, the simulation was initiated with patchy particles distributed randomly in the simulation box, designed to meticulously observe and record the time frame leading to the first emergence of the target structure. In the latter case, the simulation was initiated with a pre-formed target structure to diligently monitor and analyze the enduring stability and structural integrity of the assembly. 
 Using these two different initial conditions, we aimed to acquire a comprehensive understanding of the dynamics and robustness of the structures formed under equilibrium and nonequilibrium conditions.
 Throughout our study, we presented the median values of $T_{fas}$ and $T_{stable}$, and showed the full distribution of the assembly and disassembly times {whenever possible}. 
 The rationale is attributed to the fact that in order to compute the mean, we must consider realizations in which the systems successfully assemble for $T_{fas}$ or disassemble for $T_{stable}$. These would be highly computationally expensive, particularly for lower interaction energies and very high interaction energies, respectively.

\subsection{Monte Carlo Simulation}

Our in-house program of patchy particles included both equilibrium and nonequilibrium MC simulations. The MC moves comprised both the translation of the beads and the rotation of patches relative to the center of the bead.
All the MC simulations in this work were performed at a temperature of $65$ K, and we have employed reflective walls. 
The simulation lengths, in terms of the number of MC steps, were set to $10\times10^6$ for realizations starting from both random and target configurations to determine $T_{fas}$ and $T_{stable}$ for the $8$ and $10$ patchy particle systems. For the $13$-patchy particle system, given its increased complexity and the intricate nature of the target structure, the number of MC steps was increased to $15\times10^6$ to adequately capture the assembly process and ensure the formation of the target structure.

\subsubsection{Equilibrium Simulations}

In equilibrium MC simulation, each step involved assessing the total interaction energy of the particles, both before and after a proposed MC step. 
We then employed the Metropolis criterion \cite{Metropolis1953}, which provided the probability for accepting or rejecting a move, according to the energy difference,  
$min\Big[1,exp\big\{-(U^{new}_{Particle}-U^{old}_{Particle})/k_{B}T\big\}\Big]$, 
where $U_{Particle}^{new}$ and $U_{Particle}^{old}$ represent the new and old total energies, $k_{B}$ is the Boltzmann constant and $T$ is the temperature.

\subsubsection{Nonequilibrium Simulations}

In our nonequilibrium MC simulation, we introduced an external driving force, denoted as $\epsilon_{drive}$, to the patches of the particles. 
This external force disrupted the equilibrium state of the system by deviating from the detailed balance condition, significantly changing the interaction dynamics between the particles. 
We applied this bias in a targeted manner, depending on whether the patches were moving towards or away from each other, aligning with the topology of the desired target structure to both form and maintain it. 

The acceptance or rejection of a proposed move was adjected to include the drive.
If two patches of two different patchy particles, which were the nearest neighbor of each other in the target structure, came into proximity in an MC step, the drive would render the proposed move more favorable by modifying the acceptance probability to $min \big[1,exp\big\{-(U^{new}_{Particle}-\epsilon_{drive}-U^{old}_{Particle}) / k_{B}T\big\}\big]$.
When two neighboring patches, which are adjacent in the target structure, move apart during an MC step, the drive would render the bond breaking less probable, by the modified acceptance probability to $min \big[1,exp\big\{-(U^{new}_{Particle}+\epsilon_{drive}-U^{old}_{Particle}) / k_{B}T\big\}\big]$.
When a virtual bond between two nearest neighbor patches, according to the target structure, persisted over successive MC steps, no external bias was added nor subtracted, and the acceptance probability followed the traditional Metropolis algorithm.
The application of an external driving force ($\epsilon_{drive}$) in nonequilibrium simulations is hypothesized to significantly enhance the efficiency of target structure formation and its subsequent stability, by selectively favoring assembly pathways that align with the desired target structure.

Our approach of external drive replicates the directed self-assembly processes found in experimental systems like DNA nanostructures and colloidal crystals \cite{Mariottini2021, Mazzotti2023, Berg2023, Byrom2013}. In these experiments, specific interactions and external stimuli, such as light, pH variations, and electromagnetic fields, guide the formation of the desired structures. For instance, light and pH control in DNA assembly and disassembly enables DNA structures to adapt to various stimuli, exhibiting properties necessary for synthetic cells \cite{Mariottini2021, Mazzotti2023, Berg2023}. Similarly, applying magnetic fields in colloidal crystal assembly selectively stabilizes favorable configurations by adjusting energy levels \cite{Byrom2013}. These experimental analogs ensure that our simulations are grounded in realistic and practically achievable conditions.

Complementary to the $8$, $10$, and $13$ patchy particle simulations aimed to understand the formation and stability of the target structures, comparative analyses were conducted to assess the effects of patch length variation and simulation cell dimension variation.
These were carried out in $10$ patchy particle systems to achieve insights into the factors influencing the self-assembly.

\subsection{Molecular Dynamics Simulations}

\subsubsection{Equilibrium Simulations}

Equilibrium MD simulations were conducted using the Large-scale Atomic/Molecular Massively Parallel Simulator (LAMMPS) software \cite{Thompson2022} within the canonical ensemble.
We focused on two specific systems comprising $8$ and $10$-patchy particles, each featuring only a single internal state.
Consistent with the MC simulations, the dimensions of the simulation cell for these systems were maintained at $8$ and $9$ \AA, respectively, and periodic boundary conditions (PBC) were employed in all directions.
The target structure for these simulations, depicted in Figs. \ref{target}(a) and (b), maintained uniform colors for all beads across both systems according to the single state particles.

Following the study of Jover \emph{et.~al.}, the bead of the patchy particle is modeled with pseudo-hard-sphere (PHS) potential \cite{Jover2012}:
\begin{equation}
\label{bead}
    U_{bead} =
    \begin{cases}
        \lambda_{r}\big(\frac{\lambda_{r}}{\lambda_{a}}\big)^{\lambda_{a}}\epsilon_{r}\Big[\big( \frac{\sigma_{bead}}{r} \big)^{\lambda_{r}} - \big( \frac{\sigma_{bead}}{r} \big)^{\lambda_{a}} \Big]+\epsilon_{r}, \ \  \text{if} \ r < \big(\frac{\lambda_{r}}{\lambda_{a}} \big)\sigma_{bead}\\
        0 , \ \ \ \ \ \ \ \ \ \ \ \ \  \ \ \ \ \ \ \ \ \ \ \ \ \  \ \ \ \ \ \ \ \ \ \ \ \ \  \ \ \ \ \ \ \ \ \ \  \text{if} \  r \geq \big(\frac{\lambda_{r}}{\lambda_{a}} \big)\sigma_{bead}
    \end{cases}
\end{equation}
where $\lambda_{a} = 49$ and $\lambda_{r} = 50$ are the coefficients for attractive and repulsive forces, respectively.
The parameter $\epsilon_{r}$ represents the interaction strength in the PHS model, characterizing the energy associated with the interactions.
The variable $r$ signifies the distance measured from the center of one particle to another.
Further, the values for $\epsilon_{r}$ and $\sigma_{\text{bead}}$ are set to be $0.24$ kJ/mol and $2$ \AA, respectively, aligned with the parameters utilized by Espinosa \emph{et al.} in their investigation of liquid-vapor coexistence in systems comprising patchy particles \cite{Espinosa2019}. These parameters have been established to provide a realistic representation of particle interactions within the simulated environment.

The interaction energy between patches of different particles, $U_{patch}^{csw}$, is modeled with a continuous attractive square-well potential\cite{Espinosa2014}:
\begin{equation}
\label{csw}
    U_{patch}^{csw} = - \frac{1}{2} \epsilon_{patch} \Big[1-\tanh\Big(\frac{r-r_{w}}{\eta}\Big) \Big]
\end{equation}
where $r$ refers to the distance between the centers of two patches on different particles, 
$r_{w}$ is the radius of the attractive well,  
while the steepness of the well is controlled by $\eta$. 
In this context, $r_{w}$ is equivalent to the width of the patches previously introduced in the simulation.
In MD simulations, we have chosen the masses of the interaction sites to be 10$\%$ of the mass of the bead particle.
The specific values of the core particle and patch were taken as $10$ amu and $1$ amu.
However, the choice of masses might affect the quantitative nature but does not affect the qualitative nature of the key findings of our study since our external drive protocol does not depend on inertia.
To ensure that only one bond forms per patch, we have set the values of $\eta$ and $r_{w}$ to be $0.005\sigma_{bead}$ and $0.12\sigma_{bead}$ respectively. 

MD simulations were performed at a temperature of $40$ K using the Nose-Hoover thermostat \cite{Nose1984, Hoover1985}, which is lower compared to the $65$ K set in MC, to minimize fluctuations in the system.
Nevertheless, employing two different temperatures does not affect our conclusions, as the system dynamics inherently depend on the ratio of  $\epsilon_{patch}$  to $k_{B}T$. 
 The simulation duration for studying the self-assembly formation from both random initial conditions and starting at the target configuration was set to $12000$ ps.
We integrated the equation of motion using the velocity Verlet algorithm \cite{VelocityVerlet}  with a carefully selected timestep of $0.2$ fs, chosen to ensure the stability and non-divergence of the simulations. 
During these simulations, we recorded the trajectories and total interaction energy between patches every $100$ fs.

\subsubsection{Nonequilibrium Simulations}

In our nonequilibrium MD simulations using LAMMPS, we have integrated an external drive of a square wave potential $U_{square}(t)$. Drawing inspiration from various biological systems, such a drive resembles the voltage-gated ion channels found in neurons that open and close in response to voltage changes \cite{Terlau1998}, allowing for quick transmission of electrical signals. It is also similar to the bistable genetic switches that enable cellular phase transitions under specific conditions \cite{Fang2018}. Recent experimental studies, such as the one conducted by Song \emph{et al.} \cite{Song2015} and the capability of electromagnetic fields to direct the assembly of non-spherical particles studied by Shields \emph{et al.} \cite{Shields2013}, further support our approach.

To simulate the influence of an external periodic drive, a square wave potential, $U_{square}(t)$, is employed to modulate patchy particle interactions within the LAMMPS framework \cite{Thompson2022}, defined as:
\begin{equation}
U_{square}(t) = 
\begin{cases} 
U_{high}, & 0 \leq t < \frac{T}{2} \\
U_{low}, & \frac{T}{2} \leq t < T
\end{cases}
\label{square}
\end{equation}
where $U_{high}$ and $U_{low}$ correspond to the high energy and low energy amplitudes, and $T$ is the period of the oscillation. 
This approach simulates the effect of an external drive by periodically shifting the interaction energy between the patchy particles, thus driving the system away from equilibrium.

By adding the square wave potential to the base patchy interaction potential, $U_{patch}^{csw}$ in Eq.~\ref{csw}, we introduce a time-dependent interaction potential that oscillates between two defined energy states. These states, corresponding to the high and low energy phases of the square wave, simulate the presence of an external driving force acting on the particles (see Eqs.~$11$ and $12$ of the main manuscript). This method provides a computationally feasible technique to study the dynamic self-assembly processes under nonequilibrium conditions.

The interaction potential between two patchy particles now includes two types of terms, a spatial-dependent term, $U_{patch}^{csw}(r)$, and a time-dependent term, $U_{square}(t)$ (see Eq.~\ref{square}). 
The new modified interaction potential, $U_{patch}^{mod}(r,t)$, is given by:
\begin{equation}
U^{mod}_{patch}(r, t) = U^{csw}_{patch}(r) + U_{square}(t)
\end{equation}
and the modified total interaction energy of a patchy particle is:
\begin{equation}
    U^{mod}_{particle}(r, t) = U_{bead} (r) +  U^{mod}_{patch}(r,t)
\end{equation}
where $U^{mod}_{patch}(r, t)$ oscillates around the intrinsic patchy interaction potential, $U^{csw}_{patch}(r)$.
This oscillation introduces nonequilibrium conditions by periodically altering the interaction landscape.

\begin{figure}[ht]
    \centering
    \includegraphics[width=16cm]{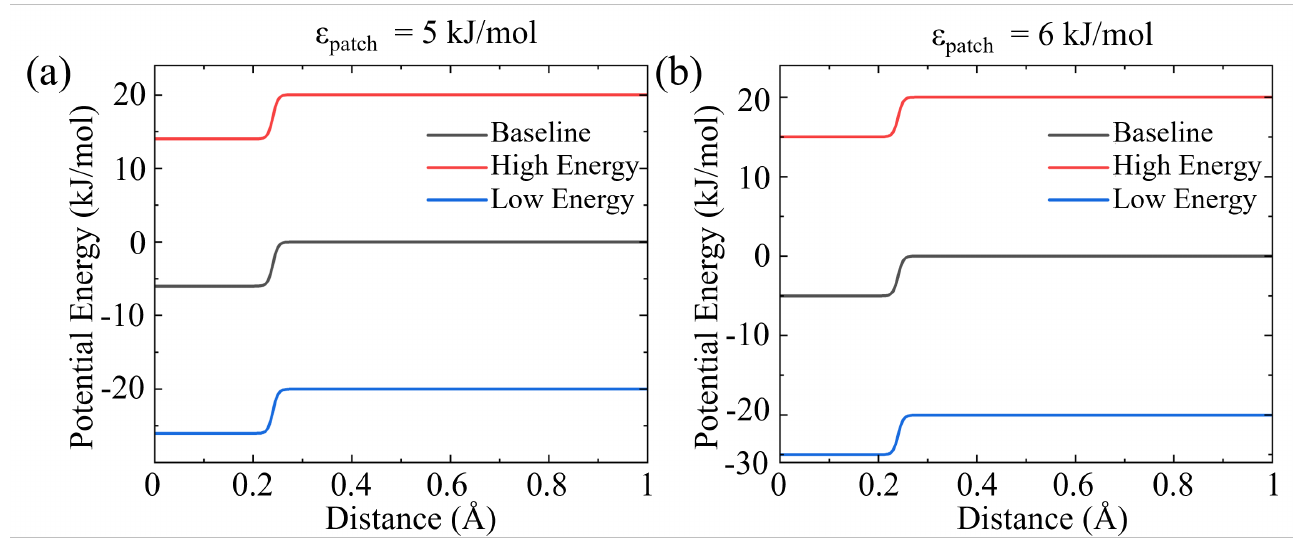}
    \caption{ Patchy interaction potential, $\epsilon_{patch}$, having a baseline value (black) of (a) $5$~kJ/mol and (b) $6$~kJ/mol, and the corresponding high energy (red) and low energy (blue) phases of the square wave potential with a potential of $20$~kJ/mol. }
    \label{patchy_poten}
\end{figure}

Fig.~\ref{patchy_poten} provides a visual representation of the square wave potential modulation of the interaction potential between patchy particles for baseline interaction energy of $5$ kJ/mol (Fig.~\ref{patchy_poten}(a)), and $6$ kJ/mol (Fig.~\ref{patchy_poten}(b)), with their respective high and low energy phases and an external wave potential amplitude of $20$~kJ/mol.

The period of the square wave, set to $20$ ps, ensures a rapid effect relative to the bond formation timescale, with amplitude variations directly mimicking the external drive adjustments in the nonequilibrium MC simulations.
These settings allow us to explore the dynamics of self-assembly and structural stability under nonequilibrium conditions through $T_{fas}$ and $T_{stable}$.
Excluding the {periodic square} potential, the rest of the simulation details, including the time-step, frequency of storing trajectories, and boundary conditions, were similar to the case of the MD simulation under equilibrium conditions.

In both MC and MD simulation, all the values of $T_{fas}$ and $T_{stable}$ were obtained from $20$ distinct simulations for a given $\epsilon_{patch}$ and $\epsilon_{drive}$ values. Randomization of the $20$ MC simulations stems from the different values of the random seed generator, whereas for MD simulations, randomization stems from different velocity distributions from the random seed generator.

We also simulate both MC and MD for a larger system comprising $100$ particles in a simulation box with dimensions of $30 \times 30 \times 30$ \AA$^{3}$, maintaining a number density of $0.003$ \AA$^{-3}$. This is consistent with our previous simulation of $13$ patchy particles in a cubic box with a side length of $15$ \AA, using a single interaction state between the patches. The simulation cell walls are modeled with PBC. The duration of the MC simulations is $5 \times 10^{6}$ MC steps, and the MD simulation is $10$ ns. During the simulation, we maintain a neighbor list with a cutoff distance of $10$ \AA, updating this list every $10$ MC steps for MC simulation and $2$ fs for MD simulation. The particles interact with each other through PHS potential (Eq.~\ref{bead}), and patches are modeled with angles set to match an $8$-ring structure and a $4$-ring structure while keeping other parameters identical to those used in the smaller-scale systems. For this large system, both MC and MD simulations were carried out at $40$ K.

\section{Results and Discussion}

\subsection{Monte Carlo Simulation Results}

\begin{figure}[ht]
   \centering
   \includegraphics[width=16cm]{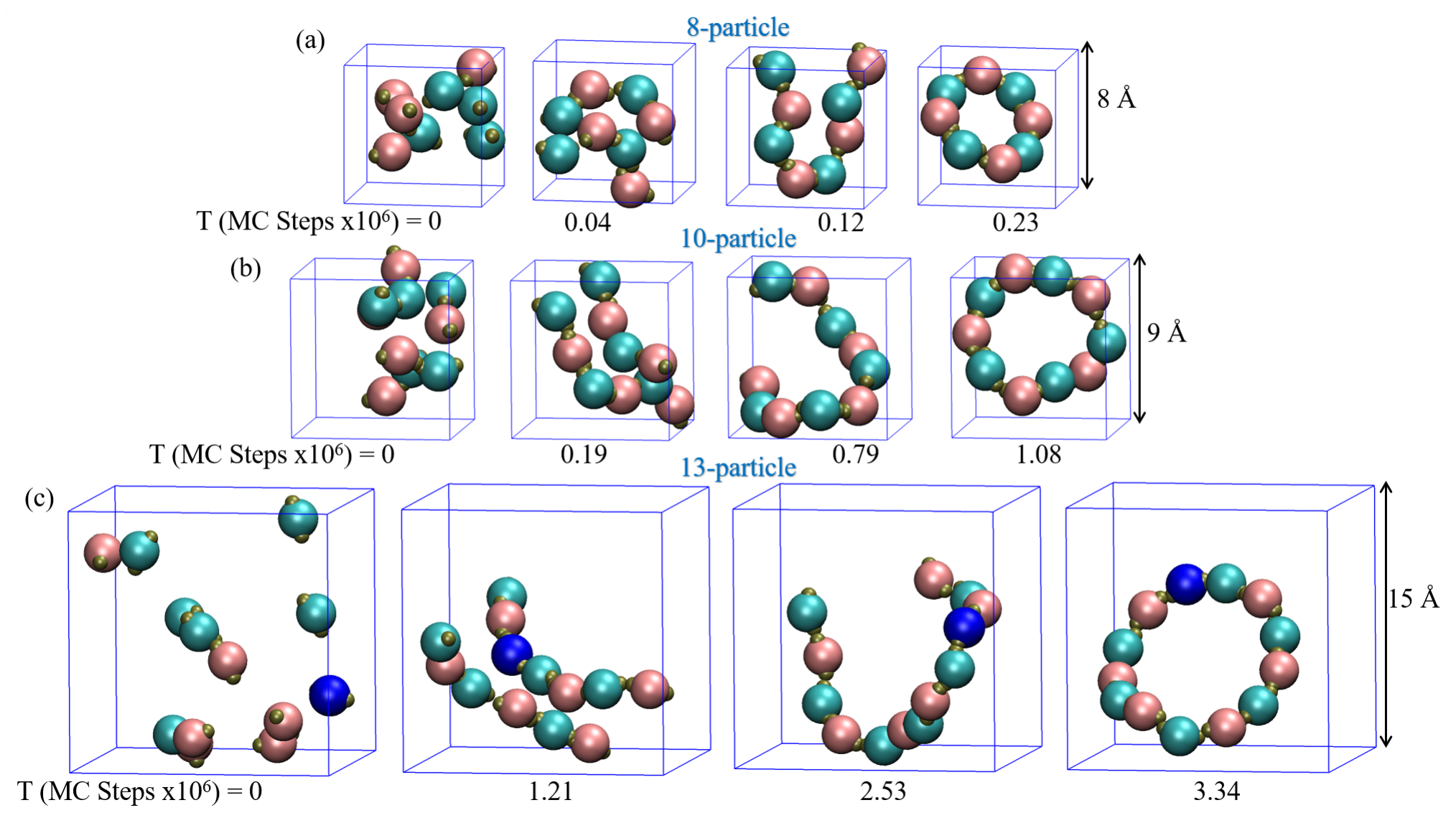}
   \caption{Equilibrium Monte Carlo simulation snapshots of (a) $8$, (b) $10$, and (c) $13$ particles of type $\alpha$ (pink), $\beta$ (cyan), and $\gamma$ (blue), shown at various MC steps, starting from random initial conditions and ending at the assembly of the ring target structure. Sizes of the simulation boxes are shown.
   }
   \label{Snapshot}
\end{figure}

In our study, we begin by conducting equilibrium Monte Carlo simulations for three different types of systems. The first two systems consist of $8$ and $10$ patchy particles with two internal states denoted by $\alpha$ and $\beta$. The third system consists of $13$ patchy particles with three internal states denoted by $\alpha$, $\beta$, and $\gamma$.
Each realization is initiated at random configuration, aiming to self-assemble into the desired polygonal ring structure respective to its particle count. The assembly process is documented in snapshots (Fig.~\ref{Snapshot}), from the initial disorder to the final ordered state.

For the $8$-particle system (Fig.~\ref{Snapshot}(a)), snapshots at initial, intermediate ($0.04\times10^{6}$ and $0.12\times10^{6}$ MC steps), and final stages ($0.23\times10^{6}$ MC steps) illustrate a straightforward path to the target structure, indicating a simple energy landscape conducive to rapid assembly.
The $10$-particle system (Fig.~\ref{Snapshot}(b)) showcases a more gradual formation process. Initial and subsequent snapshots taken at $0.19\times10^{6}$, $0.79\times10^{6}$, and $1.08\times10^{6}$ MC steps reveal an extended assembly time influenced by the larger particle count, underscoring the scaling challenges in self-assembly processes.
The assembly of a $13$-particle system (Fig.~\ref{Snapshot}(c)) is the most intricate, achieving the target state at $3.34\times10^{6}$ MC steps after navigating through various stages.
These snapshots collectively display the inherent complexity of each system and exemplify the typical pathway to the target configuration.

These configurations, captured at different MC steps and conducted under an interaction energy of $10$ kJ/mol, highlight the kinetic pathways and the balance between entropic and energetic factors influencing assembly. As the number of particles increases, so does the complexity of the energy landscape, leading to longer assembly times. 
This emphasizes the potential for more intermediate metastable states in larger systems and the critical role of interaction energy in assembly efficiency. 
Next, we set to explore how the time to first assembly ($T_{fas}$) and stability ($T_{stable}$) depend on the patchy interaction potential ($\epsilon_{patch}$) across different system sizes.

\subsubsection{Time to First Assembly in Equilibrium MC Simulations}
\label{Tfas_equi_MC}

In an attempt to calculate the time to reach the first assembly from random configurations to understand the inherent trade-off of equilibrium simulations, we carry out MC simulation over a wide range of $\epsilon_{patch}$ values, ranging from $0$ to $50$ kJ/mol. 
A temperature of $65$ K corresponds to $0.54$ kJ/mol ($k_{B}T$). Therefore, the range of $\epsilon_{patch}$, between $0$ to $50$ kJ/mol, translates to a range of $0$ to $93$ $(\epsilon / k_{B}T)$.
For each value of $\epsilon_{patch}$, the value of the time to first assembly, $T_{fas}$, is quantized in MC steps.

\begin{figure}[!h]
    \centering
    \includegraphics[width=16cm]{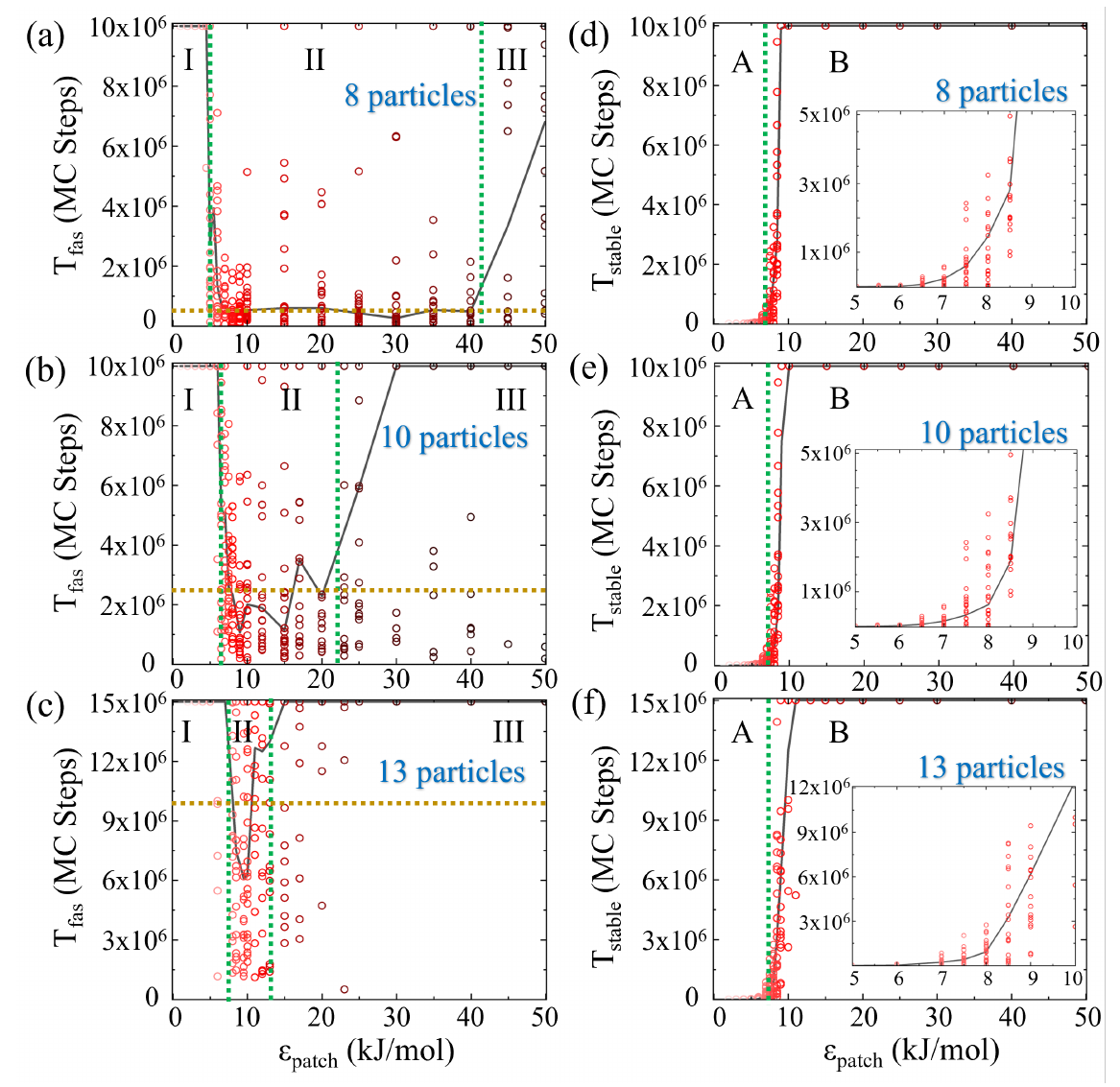}
    \caption{Equilibrium MC simulation results. 
    Median $T_{fas}$ values (black line) of $20$ individual realizations (red circles), across different $\epsilon_{patch}$ values, for (a) $8$ patchy particle system, (b) $10$ patchy particle system, and (c) $13$ patchy particle system. 
    The $\epsilon_{patch}$ range is segmented into three regions (green dotted lines), with the average $T_{fas}$ over the median values in the middle Region II depicted by a golden dotted line. 
    Median $T_{stable}$ values across different $\epsilon_{patch}$ values for (d) $8$ patchy particle system, (e) $10$ patchy particle system, and (f) $13$ patchy particle system. 
    The $\epsilon_{patch}$ range is categorized into two distinctive zones (green dotted lines).
    Insets: Zoom into the lower $\epsilon_{patch}$ range.}
    \label{Tfas_Tstable_equilibrium}
\end{figure}

For the $8$ particle (Fig.\ref{Tfas_Tstable_equilibrium}(a)), $10$ particle (Fig.\ref{Tfas_Tstable_equilibrium}(b)), and $13$ particle (Fig.~\ref{Tfas_Tstable_equilibrium}(c)) systems, the median value of $T_{fas}$ remains relatively constant at the simulation length ($10\times10^{6}$ MC steps) at low interaction energies ($\epsilon_{patch}$), suggesting that particles do not coalesce into a stable assembly within the allotted simulation time, with $\epsilon_{patch}$ ranges of $0$ to $5$ kJ/mol, $7$ kJ/mol, and $8$ kJ/mol, respectively.
These low $\epsilon_{patch}$ ranges which hinder assembly formation are referred to as Region I (see corresponding  Movies~S$1$, S$2$, and S$3$, for $8$, $10$, and $13$ patchy particles in the SI).

As $\epsilon_{patch}$ increases and falls within and intermediate range, a notable decrease is observed in $T_{fas}$. These ranges are $6$ -- $40$~kJ/mol for the $8$ particle system, $8$ -- $20$~kJ/mol for the $10$ particle system, and $8$ -- $12$~kJ/mol for the $13$-particle system.
This decrease stabilizes to  $0.7\times10^{6}$, $2.1\times10^{6}$, and $9.7\times10^{6}$  MC steps for the $8$, $10$, and $13$ patchy particle systems, respectively, as indicated by the golden dotted lines in Figs.~\ref{Tfas_Tstable_equilibrium}(a)-(c). 
These values are obtained by averaging over the median values of $T_{fas}$ in the intermediate Region II.
The observed narrowing of this optimal interaction energy range with increasing system complexity from $8$ to $13$ patchy particles can be attributed to the intricate potential energy landscapes in larger assemblies.
As the number of particles grows, the likelihood of encountering metastable states increases, necessitating a more precise tuning of interaction energies to navigate toward the desired assembly. 
This intermediate region (designated as Region II) is identified as the optimal energy regime for assembly at equilibrium, where the majority of simulations successfully reach the target structure, albeit with an increase in $T_{fas}$.

At higher $\epsilon_{patch}$ values, $T_{fas}$ begins to rise in all of the systems, suggesting that excessively strong interactions may be counterproductive, potentially trapping the system in kinetic traps.
Significantly, for the $10$ and $13$ patchy particle systems, the median value of $T_{fas}$ approaches the total number of MC steps, indicating more than half of the simulations fail to reach the target structure. 

These trends across different interaction energy regimes highlight the delicate balance required for the self-assembly of patchy particles. 
Such insights are pivotal in advancing the fabrication of nanomaterials and the understanding of biological self-assembly processes where precise control over interaction energies is often beyond reach.

\subsubsection{Target Stability in Equilibrium MC Simulations}
\label{Tstable_equi_MC}

The target stability is represented by the time the system remains in the target structure, $T_{stable}$. It is shown for systems comprising $8$, $10$, and $13$ patchy particles in Figs.~\ref{Tfas_Tstable_equilibrium}(d), \ref{Tfas_Tstable_equilibrium}(e), and \ref{Tfas_Tstable_equilibrium}(f), respectively. At low $\epsilon_{patch}$ values (Region A), all systems display relatively low $T_{stable}$, indicating insufficient stability of the target structures. As $\epsilon_{patch}$ increases to values larger than $\sim 6-7$ kJ/mol (Region B), $T_{stable}$ rises, indicating that the structures remain assembled throughout the entire simulation. For the $8$-particle system, there is a sharp increase in $T_{stable}$ beyond the critical threshold, while the $10$ and $13$-particle systems show a more gradual transition.

\subsubsection{Analysis of Assembly Time and Stability in Equilibrium Self-Assembly}
\label{compara_MC}

The dynamics of assembly and stability across the $8$, $10$, and $13$ patchy particle systems under equilibrium conditions are influenced by interaction energy, $\epsilon_{patch}$, as evident by the median values we have plotted. Plotting median values is necessary here due to the fact that in some regions of the parameter space, there were many realizations in which the target was not assembled (for $T_{fas}$) or disassembled (for $T_{stable}$), particularly in the case of low and high interaction energy values, which would bias the mean values. Therefore, showing the median ensures that the derived insights accurately reflect the typical system behavior. 
Low interaction energies prove insufficient for both assembly and stability across all systems, as indicated by the high values of $T_{fas}$ (Region I) and low values of $T_{stable}$ (Region A).
At intermediate interaction energies, a clear decrease in $T_{fas}$ (Region II) highlights the optimal range for self-assembly under equilibrium conditions. This range becomes narrower as system complexity increases from $8$ to $10$ and $13$ patchy particles.  
In terms of target stability, beyond a $\epsilon_{patch}$ threshold, the assembled structure remains at the target state (Region B). 
Both $T_{fas}$ and $T_{stable}$ suggest a critical $\epsilon_{patch}$ range that balances the interaction energy necessary for assembly and subsequent stability, with larger systems requiring finer energy tuning to achieve and maintain assembled structures.
At high interaction energies, an increase in $T_{fas}$ (Region III) suggests that overly strong interactions may lead to kinetic traps and impede target structure formation. 
Concurrently, $T_{stable}$ maintains its value equal to the simulation length for the higher values of $\epsilon_{patch}$.

In the realm of equilibrium self-assembly, the behaviors within Regions I and A, characterized by longer assembly times and low target stability, respectively, present a potential for external driving forces to play a critical role in accelerating the assembly while improving the stability of the targets. 
Such an intervention is particularly crucial in these regions, as it could overcome the limitations of equilibrium self-assembly, facilitating rapid assembly and longer stability.
Furthermore, the range of interaction energy falling into the optimal region (Region II in Fig.~\ref{Tfas_Tstable_equilibrium}) for any system depends on the system complexity, including its size, the nature of interactions, the parameters of patches, and the number density.

\subsubsection{Time to First Assembly in Nonequilibrium MC Simulations}
\label{Tfas_nonequi_sec}

Nonequilibrium intervention in the form of an external drive has the potential to positively influence both the assembly speed and the target stability in Regions I and A, respectively. Here, we implement a driving force that favors the bond formation of nearest neighbor particles according to the target structure and disfavors bond breaking in this case.

\begin{figure}[!ht]
    \centering
    \includegraphics[width=16cm]{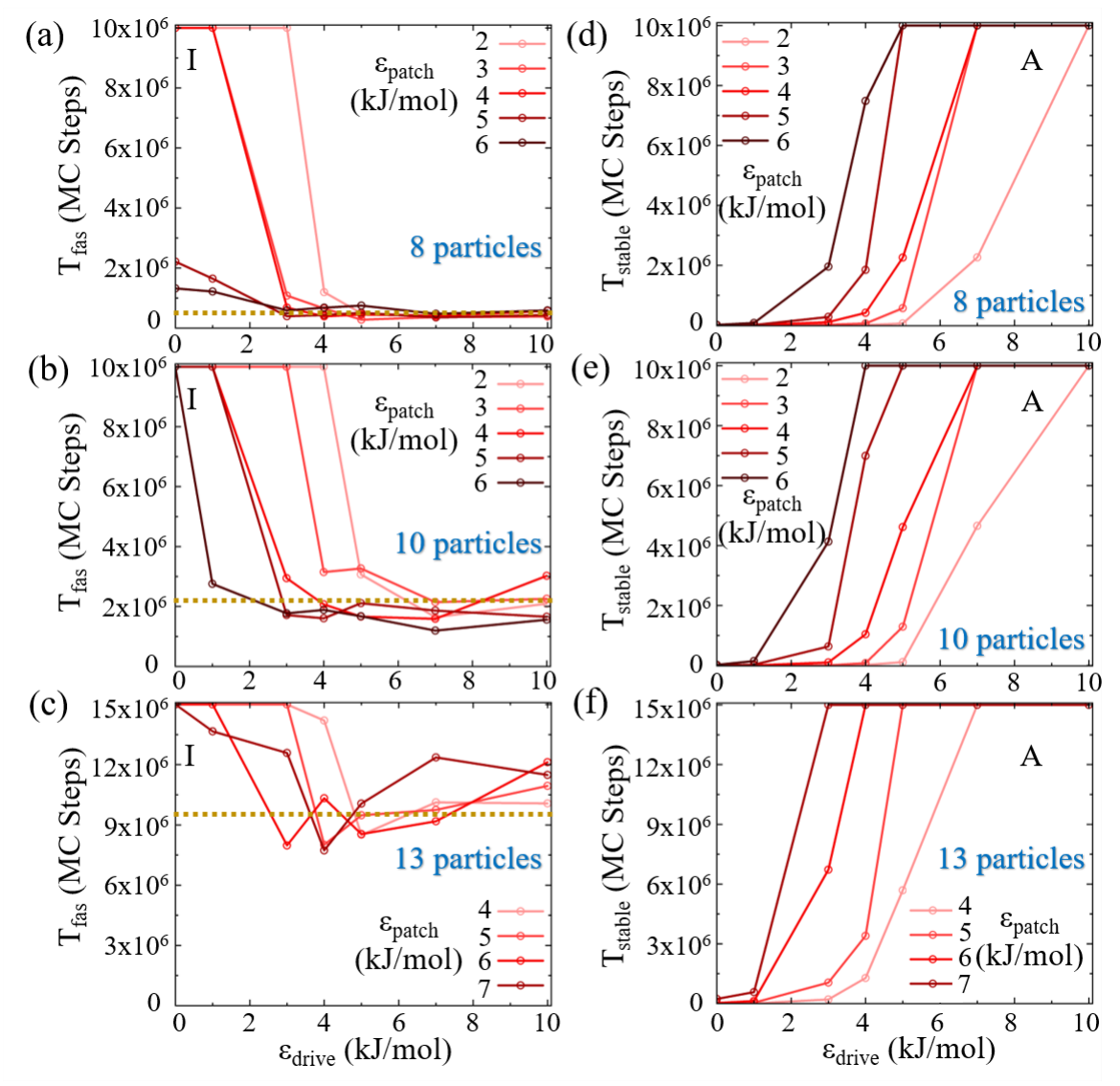}
    \caption{Nonequilibrium MC simulation results. Median $T_{fas}$ values of $20$ individual realizations for different patchy particle interaction energies, $\epsilon_{patch}$, as a function of the drive value, $\epsilon_{drive}$, for (a) $8$ particle system, (b) $10$ particle system, and (c) $13$ particle system.
    The golden dotted line denotes the average value of $T_{fas}$ in the intermediate Region II obtained from the corresponding results presented in Fig.~\ref{Tfas_Tstable_equilibrium}(a)-(c).
    Median $T_{stable}$ of $20$ individual realizations for different patchy particle interaction energies, $\epsilon_{patch}$, as a function of the drive value, $\epsilon_{drive}$, for (d) $8$ particle system, (e) $10$ particle system, and (f) $13$ particle system.
    }
    \label{Tfas_Tstable_nonequilibrium}
\end{figure}

In the $8$-particle system (Fig.~\ref{Tfas_Tstable_nonequilibrium}(a)) for $\epsilon_{patch}$ of $2$~kJ/mol, $T_{fas}$ remains at the maximum simulation length until $\epsilon_{drive}$ exceeds $3$ kJ/mol, indicating that the system does not achieve the target structure without an adequate driving force. When $\epsilon_{drive}$ reaches $4$~kJ/mol, $T_{fas}$ is reduced significantly, suggesting that a specific threshold of an external drive is required to facilitate assembly.
For $\epsilon_{patch}$ values of $3$ and $4$~kJ/mol, $T_{fas}$ decreases to the equilibrium value of Region II ($0.7\times10^{6}$ MC steps) once $\epsilon_{drive}$ exceeds $3$ kJ/mol.
A higher value of $\epsilon_{patch}$ ($5$-$6$ kJ/mol) leads to a relatively small $T_{fas}$ even without an external drive, highlighting the importance of intrinsic interaction energy in assembly efficiency.
In the later cases, a minimal $\epsilon_{drive}$ of $1$ kJ/mol is adequate to reduce $T_{fas}$ to its lowest observed plateau.
To achieve similar reductions in $T_{fas}$ for the $10$-particle system, converging to the average $T_{fas}$ value of the corresponding Region II (Fig.~\ref{Tfas_Tstable_equilibrium}(b)), a higher $\epsilon_{drive}$ of $5$~kJ/mol is required for $\epsilon_{patch}$ value of $2$~kJ/mol.
As $\epsilon_{patch}$ increases from $3$ to $6$~kJ/mol, lower $\epsilon_{drive}$ is required.

For the $13$-particle system (Fig.~\ref{Tfas_Tstable_nonequilibrium}(c)), $T_{fas}$ remains at the simulation length value ($15\times10^{6}$) without the intervention of external drive for all $\epsilon_{patch}$ values. 
An external drive of $4$~kJ/mol is needed at $\epsilon_{patch} = 4$ kJ/mol to reach the target, with higher $\epsilon_{drive}$ reducing $T_{fas}$ to the equilibrium average value of Region II (Fig.~\ref{Tfas_Tstable_equilibrium}(c)). As $\epsilon_{patch}$ increases from $5$ to $7$ kJ/mol, lower $\epsilon_{drive}$ values are needed.

Comparing the various system sizes, for $\epsilon_{patch}$ value fixed at $6$~kJ/mol, a drive value $\epsilon_{drive}$ of $3$~kJ/mol is required to reduce $T_{fas}$ to its equilibrium value in the $13$ particle system, where only $1$~kJ/mol is required to achieve the same result in a $10$ particle system, and the $8$ particle system archives the equilibrium value without any external drive.

These simulations show that external drives can effectively reduce the assembly times up to a point where an increase in the driving force does not result in additional acceleration. This suggests that while external driving forces can enhance assembly speed, there exists a limit to their efficacy beyond which the system behaves similarly to the equilibrium systems observed in the corresponding Region II. 
For a visualization of the nonequilibrium self-assembly process, see Movies~S$4$, S$5$, and S$6$, for $8$, $10$, and $13$ patchy particles, respectively, in the SI.

\subsubsection{Target Stability in Nonequilibrium MC Simulations}

The external drive also increases the target stability, manifested in higher $T_{stable}$ for increasing drive values for several $\epsilon_{patch}$ values within the corresponding Region A for all systems tested, of $8$ (Fig.~\ref{Tfas_Tstable_nonequilibrium}(d)), $10$ (Fig.~\ref{Tfas_Tstable_nonequilibrium}(e)), and $13$ (Fig.~\ref{Tfas_Tstable_nonequilibrium}(f)) patchy particles, aligning with previous findings \cite{Bisker2018, Ben2021, Faran2023}.

In the $8$-particle system (Fig.~\ref{Tfas_Tstable_nonequilibrium}(d)), at $\epsilon_{patch} = 2$ kJ/mol, $T_{stable}$ gradually increases with $\epsilon_{drive}$, aligning with the simulation length ($10\times10^{6}$ MC steps) beyond $5$ kJ/mol. As $\epsilon_{patch}$ increases from $3$ to $6$ kJ/mol, lower $\epsilon_{drive}$ values suffice to enhance $T_{stable}$, indicating higher internal interaction energies improve stability. For the $10$-particle system (Fig.\ref{Tfas_Tstable_nonequilibrium}(e)), $T_{stable}$ at $\epsilon_{patch} = 2$ kJ/mol increases only beyond $\epsilon_{drive} = 6$~kJ/mol, with higher $\epsilon_{patch}$ reducing the needed $\epsilon_{drive}$. In the $13$-particle system (Fig.\ref{Tfas_Tstable_nonequilibrium}(f)), at $\epsilon_{patch} = 4$~kJ/mol, $\epsilon_{drive} = 3$~kJ/mol significantly increases stability. For higher $\epsilon_{patch}$ values, modest increases in $\epsilon_{drive}$ ensure maximal stability.

\subsubsection{Analysis of Nonequlibrium Self-Assembly Mechanism}

We now turn to explore the assembly dynamics of patchy particles under external forces by focusing on an order parameter ($R$) and the total entropy production ($S$) \cite{Seifert2012} as a function of MC steps. 
$R$ is defined as the ratio of the number of formed bonds at any given MC step to the total number of bonds in the target structure, therefore quantitatively characterizes the path to assembly.
The detailed description of the computation of $S$ in our MC simulation is provided in Section S$1$ of the Supporting Information (SI).

\begin{figure}[ht]
    \centering
    \includegraphics[width=16cm]{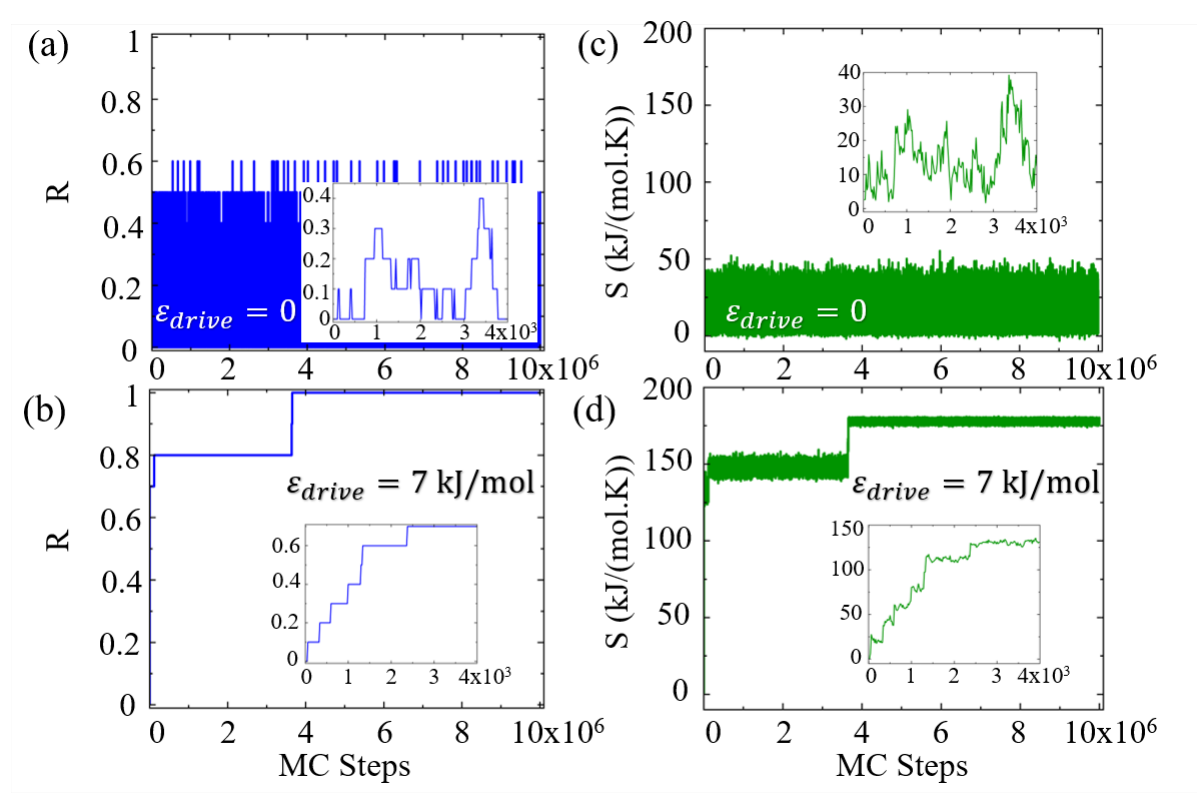}
    \caption{Order parameter ($R$) as a function of MC steps (a) in equilibrium and (b) under nonequilibrium conditions with an external drive $\epsilon_{drive}=7$~kJ/mol.
    The total entropy production ($S$) as a function of MC steps (c) in equilibrium and (d) under nonequilibrium conditions with an external drive $\epsilon_{drive}=7$~kJ/mol
    Results are presented for a single MC realization of $10$ particle system with $\epsilon_{patch}=3$~kJ/mol.
    The inset plots show Zoom in into the early simulation steps.}
    \label{thermo_prop}
\end{figure}

The evolution of $R$ for a $10$ patchy particle system with patch interaction value of $\epsilon_{patch} = 3$~kJ/mol, under equilibrium, $\epsilon_{drive} = 0$, and nonequilibrium conditions, $\epsilon_{drive} = 7$~kJ/mol, is depicted in Fig.~\ref{thermo_prop}(a) and Fig.~\ref{thermo_prop}(b), respectively.
In the absence of an external drive, $R$ displays stochastic behavior, oscillating between $0$ and $0.6$, manifesting the inability of the system to assemble the target.
With the external drive, a stark transition in the behavior of the order parameter is observed, with $R$ ascending from a disordered state directly toward the fully assembled target structure, corresponding to $R=1$ at $3.7\times10^{6}$ MC steps. The quantized changes in $R$ represent the discrete steps of the subsequent formation of $10$ bonds, ultimately yielding the $10$ particle ring structure.

Similarly, the total entropy production, $S$, exhibits fluctuations at equilibrium (Fig.~\ref{thermo_prop}(c)), indicating vanishing entropy production rate, as expected \cite{Seifert2012}.
 Under nonequilibrium conditions, the value of $S$ rapidly increases (Fig.~\ref{thermo_prop}(d)), reflecting the dissipation of energy accompanying the utilization of the external drive.

The contrast in the behavior of $R$ and $S$ with and without an external driving force emphasizes the significant role of external driving forces in directing the thermodynamic path of patchy particle systems. 
The external drive enhances the order within the system at the cost of the production of entropy \cite{Nag_Letter}.
Similar observations for trajectories of $8$ and $13$ patchy particles in equilibrium and nonequilibrium MC simulations can be found in Figs.~S$1$ and~S$2$ in Section S$2$ of the SI.

\begin{figure}[ht]
    \centering
    \includegraphics[width=16cm]{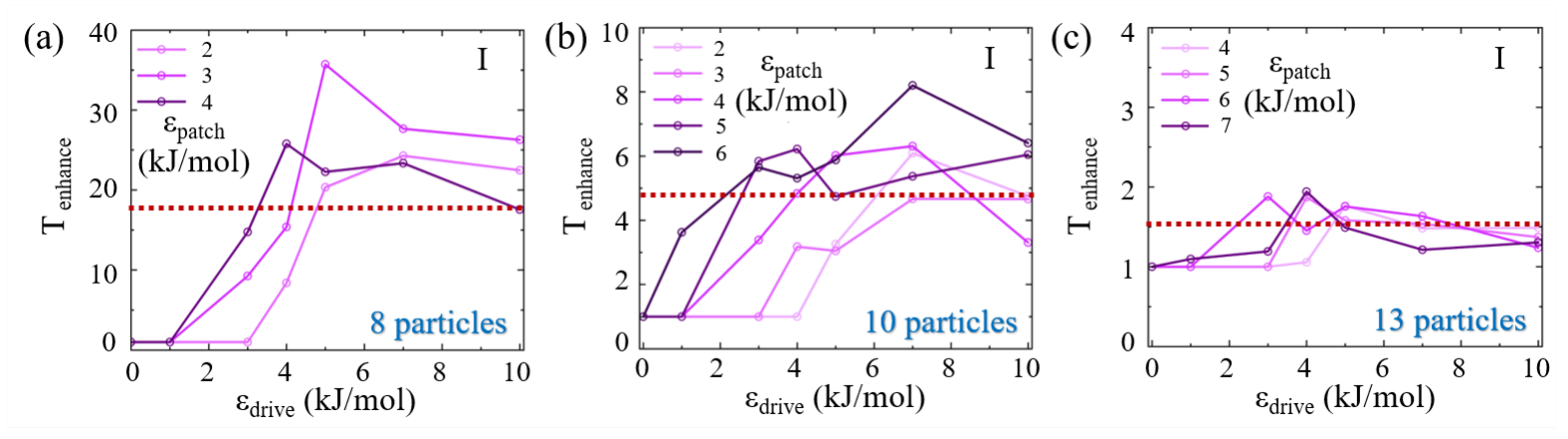}
    \caption{
    $T_{enhance}$ as a function of $\epsilon_{drive}$ for various $\epsilon_{patch}$ values from Region I for systems of (a) $8$ particle, (b) $10$ particle, and (c) $13$ particle systems.
    The brown dotted line represents the ratio of $T_{fas}$ between Regions I and II, obtained from the corresponding results in Fig.~\ref{Tfas_Tstable_equilibrium}(a)-(c).
    }
    \label{ratio_Tfas}
\end{figure}

The effect of the external drive is quantified by $T_{enhance}$, defined as the ratio of the median time to the first assembly with the external drive ($\epsilon_{drive}$) to the median time without the drive, across a set of $20$ distinct simulations for a given value of $\epsilon_{patch}$. This metric provides a measure of the efficiency improvement in assembly time attributable to the external drive. For example, this metric for $8$ particles is obtained from the values plotted in Fig.~\ref{Tfas_Tstable_nonequilibrium}(a) for a given system. 
For the $8$ particle system (Fig.~\ref{ratio_Tfas}(a)), $T_{enhance}$ shows a significant increase with increasing $\epsilon_{drive}$, showing approximately $35$ fold improvement in the assembly time compared to the equilibrium assembly for $\epsilon_{patch}=3$~kJ/mol, for example. In the $10$ particle system (Fig.~\ref{ratio_Tfas}(b)), an increase in $T_{enhance}$ with rising $\epsilon_{drive}$ is observed, with more modest improvement in assembly time. The enhancement reaches a plateau at about $6$ to $8$ fold improvement, occurring at an $\epsilon_{drive}$ near $6$ kJ/mol. 
The variation of $T_{enhance}$ of $13$ particle system is displayed in Fig.\ref{ratio_Tfas}(c), achieving a moderate improvement of up to $2$-fold in the assembly time compared to the equilibrium scenario. Compared to the increase of $T_{enhance}$ for $8$ and $10$ particle systems with the increase in drive, the increment for the $13$-particle system does not show similar improvement with increasing the drive value.
This reflects the more complex assembly pathway involving $13$ 
patchy particles and multiple interaction types: $\alpha-\beta$, $\alpha-\gamma$, and $\beta-\gamma$, compared to only $\alpha-\beta$ interaction between the other two smaller systems.
The enhancement factors $T_{enhance}$ are presented along the ratio of $T_{fas}$ from Region I to $T_{fas}$ in Region II in equilibrium simulations (Fig.~\ref{ratio_Tfas}, dotted lines), further indicating that the drive helps the system achieve its target within the time frame that matched the one of the optimal region (Region II).

In order to test the effects of our chosen patch size and particle density parameters\cite{sato2020, Chang2021}, we explore the time to first assembly and target stability for various values of the length of the patch, $\delta/2$ (Section S$3$ in the SI) and different lengths of the simulation cell (Section S$4$ in the SI).
Importantly, the trend of $T_{fas}$ and $T_{stable}$ under both equilibrium and nonequilibrium conditions remains similar, and our conclusions hold for the different cases.

\subsection{Molecular Dynamics Simulation Results}

Transitioning into the MD simulations, we expand upon the findings from the MC simulations to further explore the dynamic aspects of dissipative self-assembly of the patchy particle system under both equilibrium and nonequilibrium conditions. MD simulations, with their capacity to capture real-time dynamics, are instrumental in providing a more thorough understanding of the mechanisms at play. This approach complements the MC results by offering insights into the temporal evolution of particle assemblies, thereby enriching our analysis with a perspective that closely mirrors physical realities.

Moreover, the implementation of nonequilibrium MD simulations is a deliberate effort to bridge the computational findings with potential experimental validations. By examining the system's response to external drives and the resultant dynamic behaviors, this segment aims to not only complement the MC-derived thermodynamic landscapes but also to lay a foundational basis for future experimental investigations.

\subsubsection{Equilibrium MD Simulations}

We first set to explore the time to reach the target structure starting from random initial conditions within the simulation box for systems of $8$ and $10$ patchy particles with a single internal state, $\alpha$. The time to first assembly, $T_{fas}$, is calculated over a broad range of interaction energies between patches, $\epsilon_{patch}$, ranging from $0$ to $100$~kJ/mol. Similar to the approach taken in our equilibrium MC simulations (refer to Section~\ref{Tfas_equi_MC}), the behavior of $T_{fas}$ as a function of $\epsilon_{patch}$ can be categorized into three distinct regions. At low $\epsilon_{patch}$ values, the system is unable to achieve the target structure within the allotted simulation time of $12000$ ps due to the weak interaction (Region I), whereas for larger $\epsilon_{patch}$ values, $T_{fas}$ decreases, manifesting an optimal range for assembly (Region II). Beyond a particular $\epsilon_{patch}$ value, $T_{fas}$ begins to rise, signifying the onset of kinetic trapping that inhibits the particle reorientation towards the target structure (Region III).

\begin{figure}[!ht]
    \centering
    \includegraphics[width = 16cm]{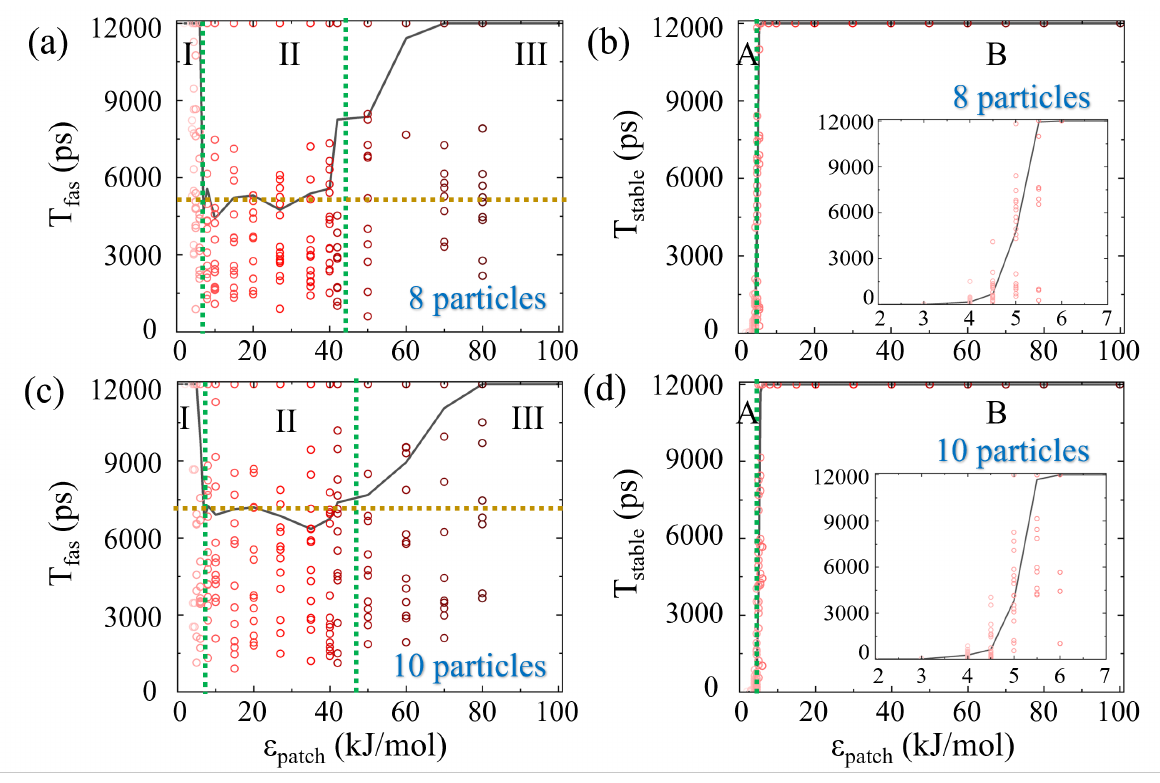}
    \caption{Equilibrium MD simulation results. 
    Median $T_{fas}$ values (black line) of $20$ individual realizations (red circles), across different $\epsilon_{patch}$ values, for (a) $8$ patchy particle system and (b) $10$ patchy particle system. 
    The $\epsilon_{patch}$ range is segmented into three regions (green dotted lines), with the average $T_{fas}$ over the median values in the middle Region II depicted by a golden dotted line. 
    Median $T_{stable}$ values across different $\epsilon_{patch}$ values for (c) $8$ patchy particle system, and (d) $10$ patchy particle system. 
    The $\epsilon_{patch}$ range is categorized into two distinctive zones (green dotted lines).
    Insets: Zoom into the lower $\epsilon_{patch}$ range. 
    }
    \label{MD_equilibrium}
\end{figure}

For the $8$ particle system (Fig.~\ref{MD_equilibrium}(a)), no assembly occurs for $\epsilon_{patch}$ values between $0$ to $6$ kJ/mol (Region I). Beyond $6$ kJ/mol, $T_{fas}$ decreases and stabilizes at $5115$ ps (Region II), until approximately $40$ kJ/mol (Region III), where $T_{fas}$ increases up to the simulation duration (Movie S$7$ in the SI). Similar results with $8$ patchy particles with hard-sphere interaction were observed in our concurrent study \cite{Nag_Letter}, where $T_{fas}$ transitioned from Region II to III above $40$ kJ/mol value of $\epsilon_{patch}$.

In addition to the time to reach the assembled state, we set to study the stability of the resulting target structure by the time the system remains at the target state, $T_{stable}$, in a set of simulations initiated at the target across the same $\epsilon_{patch}$ range. As for the MC simulations (Section~\ref{Tstable_equi_MC}), the behaviors of $T_{stable}$ clearly segment the $\epsilon_{patch}$ values into the regions, where for low $\epsilon_{patch}$ values the target quickly disassembles (Region A), whereas for higher values, the target remains stable (Region B).

In the case of the $8$ particle system (Fig.~\ref{MD_equilibrium}(b)), for $\epsilon_{patch}$ values up to $4$~kJ/mol, $T_{stable}$ is minimal (a few ps), indicating low system stability. However, as $\epsilon_{patch}$ increases, the median $T_{stable}$ rises and reaches the full duration of the simulation ($12000$ ps) by $\epsilon_{patch}=6$~kJ/mol.
These observations lead to the classification of Region A to include $\epsilon_{patch}$ values between $0$ - $6$~kJ/mol, and Region B between $\epsilon_{patch}$ values of $7$ - $100$~kJ/mol.

The same methodology is applied to a system of $10$ patchy particles with a single internal state, $\alpha$. The trend of $T_{fas}$ as a function of $\epsilon_{patch}$ mirrors the behavior of the $8$ patchy particle system (Fig.~\ref{MD_equilibrium}(c)), and the $\epsilon_{patch}$ range was similarly segmented into Region I ($0$ - $6$~kJ/mol), Region II ($7$ - $50$~kJ/mol), and Region III ($50$ - $100$~kJ/mol). The average of median $T_{fas}$ in Region II is  $7024$~ps, {shown as a golden dotted line,} which is slightly higher than that of the $8$ patchy particle system owing to its lower complexity. 
As for the stability of the resulting target quantified by $T_{stable}$ (Fig.~\ref{MD_equilibrium}(d)), we observe low target retention of only a few ps for low $\epsilon_{patch}$ values (Region A). When $\epsilon_{patch}$ exceeds $4$ kJ/mol, the median $T_{stable}$ rises and reaches the full simulation duration ($12000$ ps) at $6$ kJ/mol (Region B), (See Movie~S$8$ in the SI).

As argued in Section~\ref{compara_MC}, Regions I and A are characterized by prolonged assembly times and reduced structural stability, respectively, thereby presenting an opportunity for an external bias to overcome these equilibrium limitations and to enable quicker self-assembly and improved target stability.

\subsubsection{Nonequilibrium MD Simulation Results}
\label{nonequi_MD_sim}

In nonequilibrium MD simulations, we explore the influence of an external drive in the form of an external square wave potential with varying amplitude, $U_{square} (t)$, 
added to the patchy interaction potential, $U_{patch} (r)$, on the assembly dynamics for systems of $8$ and $10$ patchy particles. The amplitudes of $U_{square} (t)$  ranges from $0$ to $60$~kJ/mol, for patchy interaction $\epsilon_{patch}$ values of $4.5$, $5$, and $6$~kJ/mol from Regions I and A. For each scenario, we determine the time to the first assembly, $T_{fas}$, and the target stability, $T_{stable}$.

\begin{figure}[!ht]
    \centering
    \includegraphics[width=16cm]{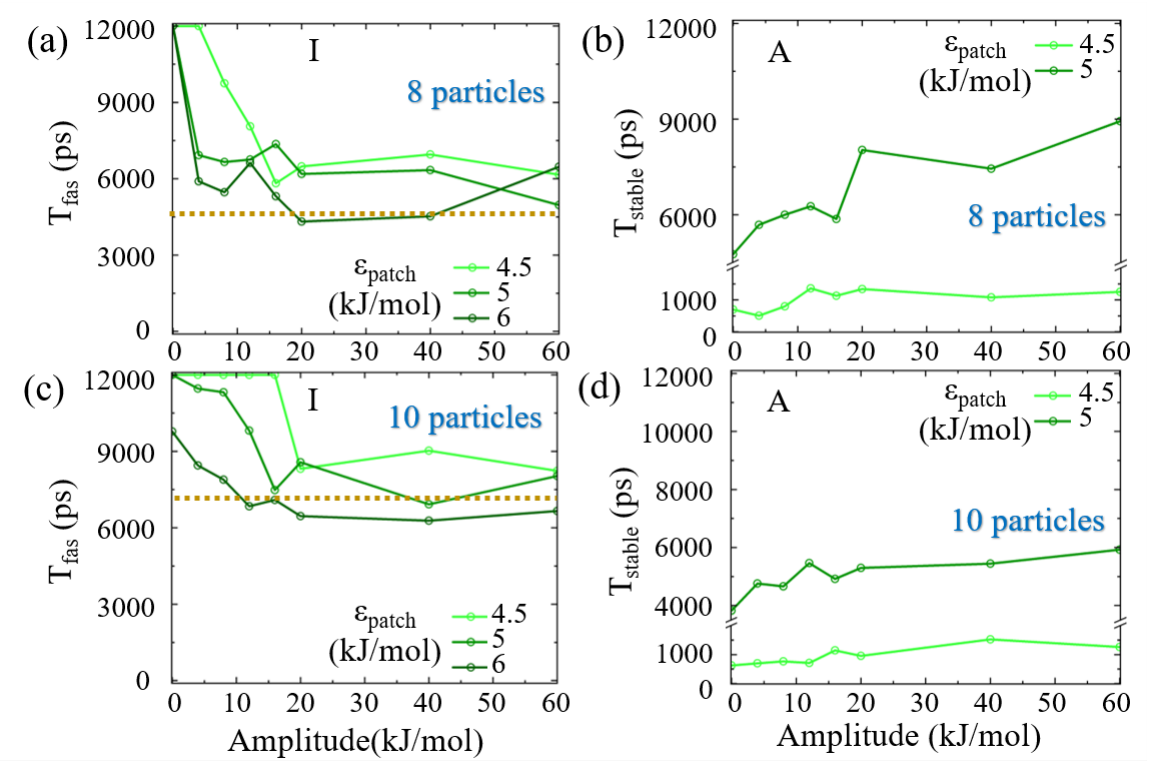}
    \caption{Nonequilibrium MD simulation results. Median $T_{fas}$ values of $20$ individual realizations for different patchy particle interaction energies, $\epsilon_{patch}$, as a function of the amplitude of the square wave potential for (a) $8$ particle system, and (c) $10$ particle system.
    The golden dotted line denotes the average value of $T_{fas}$ in the intermediate Region II obtained from the corresponding results presented in Fig.~\ref{MD_equilibrium}(a) and (b).
    Median $T_{stable}$ of $20$ individual realizations for different patchy particle interaction energies, $\epsilon_{patch}$, as a function of the amplitude of the square wave potential for (b) $8$ particle system, and (d) $10$ particle system.
    }
    \label{MD_nonequilibrium}
\end{figure}

For $8$ particle system with $\epsilon_{patch}=4.5$~kJ/mol, $T_{fas}$ significantly reduces for external drive amplitudes above $4$~kJ/mol, where for amplitude value of $20$~kJ/mol and beyond, $T_{fas}$ drops to around $6000$ ps (Fig.~\ref{MD_nonequilibrium}(a)). For $\epsilon_{patch}$ values of $5$ and $6$~kJ/mol, $T_{fas}$ immediately shows a significant decrease to around $6000$ ps (Movie S$9$ in the SI). This value is similar to the average $T_{fas}$ of approximately  $5100$ ps obtained in Region II of $8$ patchy particle equilibrium MD simulations (Fig.~\ref{MD_equilibrium}(a), golden dotted line).

In terms of stability of $8$ particle system, for $\epsilon_{patch}$ of $4.5$~kJ/mol, $T_{stable}$ is $700$ ps at zero amplitude (Fig.~\ref{MD_nonequilibrium}(b)), and increases to $1200$ ps as the amplitude of the external drive increases to $12$ kJ/mol and further. 
At $5$ kJ/mol, $T_{stable}$ rises from $4000$~ps to $9000$~ps with an increasing amplitude of up to $60$ kJ/mol. These results indicate that the introduction of an external square wave potential can substantially enhance stability for the $8$ patchy particle system.

To further complement our study, we conduct MD simulations for a $10$ patchy particle system, considering the influence of an external square wave potential.
At $\epsilon_{patch} = 4.5$ kJ/mol, $T_{fas}$ remains largely unchanged until the potential amplitude exceeds $20$ kJ/mol, then fluctuates around $7500$ ps (Fig.~\ref{MD_nonequilibrium}(c), Movie S$10$ in the SI). At $\epsilon_{patch} = 5$ kJ/mol, $T_{fas}$ decreases from $12000$~ps to $9000$~ps at $16$~kJ/mol amplitude, and further to $7000$ ps at $60$~kJ/mol. For $\epsilon_{patch} = 6$ kJ/mol, $T_{fas}$ decreases from $10000$~ps to $6000$~ps as the amplitude increases to $12$~kJ/mol, suggesting a greater sensitivity of the assembly time to the applied potential in comparison to the lower $\epsilon_{patch}$ values. 
These findings align with the average $T_{fas}$ values observed in Region II of the equilibrium MD simulations for the $10$ patchy particle system (Fig.~\ref{MD_equilibrium} (c), golden dotted line).

The stability of the $10$ patchy particle system under nonequilibrium conditions is analyzed by varying the amplitude of the external drive while initiating the system at target under $\epsilon_{patch}$ values of $4.5$ and $5$ kJ/mol (Fig.~\ref{MD_nonequilibrium}(d)). 
The value of $T_{stable}$ increases from $520$~ps to $1100$~ps as the amplitude is increased from $0$ to $16$~kJ/mol, where further increment in amplitude does not result in an additional increase in $T_{stable}$. For $\epsilon_{patch}$ value of $5$ kJ/mol, the stability time $T_{stable}$ rises from $4000$~ps to $6000$~ps with a modest amplitude increment to $10$~kJ/mol. Similarly, subsequent increases in the amplitude do not enhance $T_{stable}$ significantly.

Comparing the $8$ and $10$ patchy particle systems in nonequilibrium MD conditions, we notice distinct responses to the external drive. Both systems experience reduced assembly times ($T_{fas}$) as the amplitude of the square wave potential increases. The $10$ patchy particle system, however, shows a higher sensitivity to the amplitude changes, necessitating finer tuning in driving forces for comparable reductions in $T_{fas}$. This size-dependent response highlights how system complexity affects the dynamics of self-assembly, similar to our observation in the nonequilibrium MC simulations.
Further, stability analysis ($T_{stable}$) reveals that while the stability improves for both systems with higher drive amplitudes, the enhancement is less pronounced in the $10$ patchy particle system.
In contrast to the nonequilibrium MC simulations, where the stability ($T_{stable}$) is maintained throughout the entire simulation, the nonequilibrium MD simulations reveal further opportunities for improvement. This suggests that optimizing the effectiveness of the square wave potential as an external driving force can potentially further increase target stability.

The acceleration of the assembly and the improved target stability we find in our MD simulations are also manifested in the analysis of the order parameter, $R$, defined in a similar way as for the MC simulations, as the ratio between the number of formed bonds to the total number of bonds in the target structure. We calculate the value of $R$ averaged over the entire simulation duration and observe a {moderate} increase when the time-dependent square wave potential is introduced to the system, for systems of $8$ and $10$ particles (Section S$5$ in the SI).

In our investigation of assembly kinetics and stability under nonequilibrium conditions using MD simulations, we meticulously analyze the behavior of bond formation and dissociation events across both high and low energy phases induced by the square wave potential, $U_{square}(t)$. From this analysis, we observe that the bond durability in our patchy particle system is enhanced with the increase of square wave amplitudes, which in turn boosts the stability of the target structure and aids in its assembly (Section~S$6$ in the SI).

\subsubsection{Potential Energy and Force Comparison of $2$ Particle System}
\label{MD_2particle}

In order to better understand the dynamics of bond formation, we conduct a set of MD simulations of a system comprising only $2$ particles randomly initiated up to the first bond formation event in equilibrium and nonequilibrium conditions. During the simulation realizations, we track the total potential energy of the system and the force between the particles. For this study, the simulation cell dimensions are set to $2\times6\times6$ \AA$^3$, with patch parameters identical to those in a $10$ particle system. 
Apart from the simulation cell size, all simulation parameters are consistent with those used in simulations of the $8$ and $10$ patchy particle systems. Importantly, the outcomes do not strictly depend neither on the dimensions of the simulation cell nor on the patch parameters ($\delta/2$ and $\theta^{\text{max}}$).

\begin{figure}[ht]
    \centering
    \includegraphics[width=16cm]{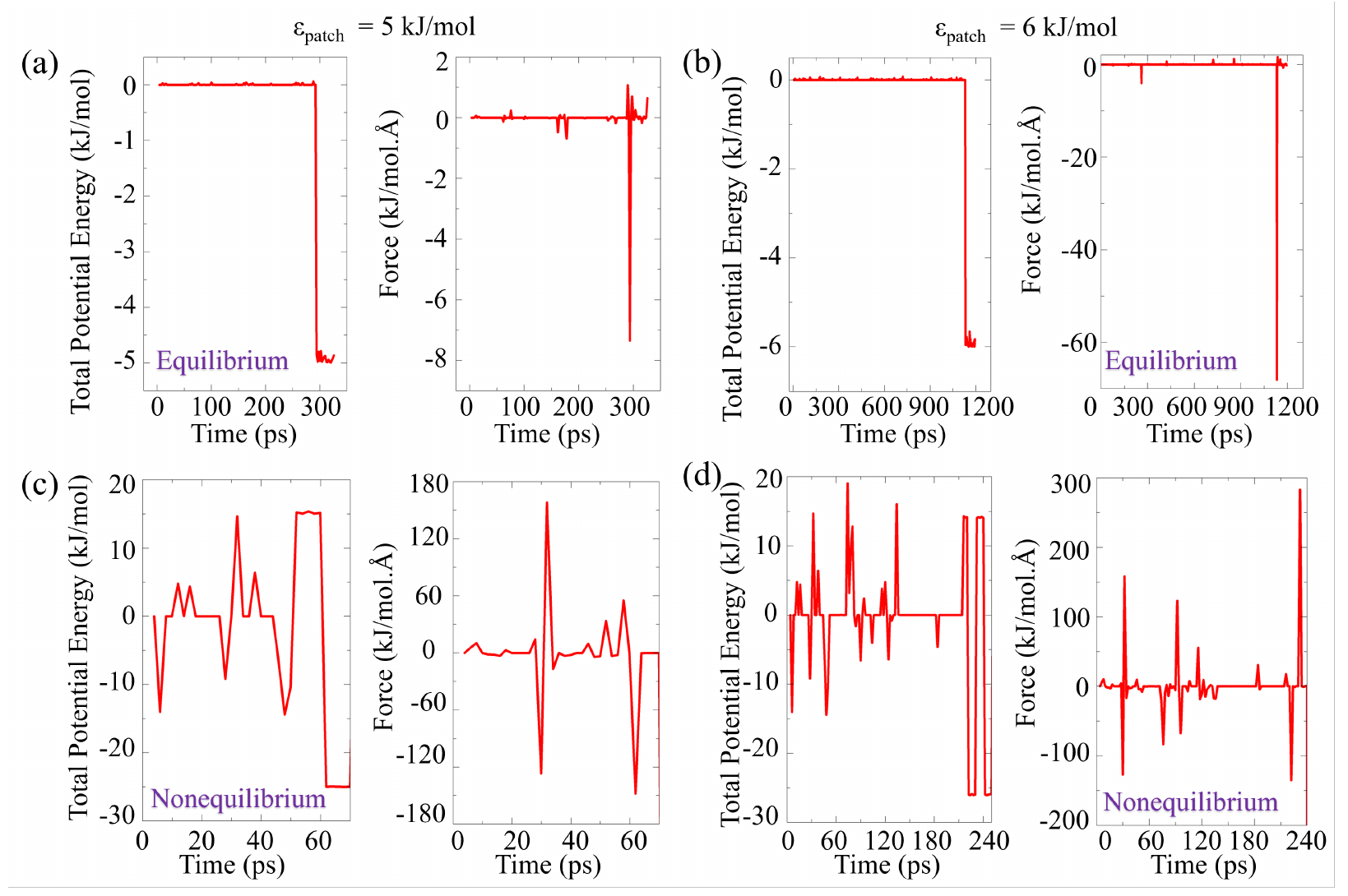}
    \caption{MD simulation of a $2$ particle system up to the first bond formation event. Total potential energy and force as a function of time for equilibrium conditions with $\epsilon_{\text{patch}}$ value of (a) $5$~kJ/mol and (b) $6$~kJ/mol, and nonequilibrium conditions with square wave potential amplitude of {$20$~kJ/mol} for $\epsilon_{\text{patch}}$ value of (c) $5$~kJ/mol and (d) $6$~kJ/mol, shown for individual realizations.
    }
    \label{MD_PE_force}
\end{figure}

Under equilibrium conditions for patchy interaction energies $\epsilon_{patch}$ of $5$~kJ/mol (Fig.~\ref{MD_PE_force}(a)) and $6$~kJ/mol (Fig.~\ref{MD_PE_force}(b)), the potential energy initially fluctuates around $0$, and the corresponding force between the particles, reflecting the gradient of the potential energy, also remains close to $0$. On the other hand, the bond formation event, which occurs at {$294$~ps and $1132$~ps for $\epsilon_{patch}$ of $5$~kJ/mol and $6$~kJ/mol}, respectively, is manifested in a noticeable decrease of the potential and a concurrent force spike.

In stark contrast, nonequilibrium conditions exhibit a completely different profile of the total potential energy and the force for both patchy interaction energies of $5$~kJ/mol (Fig.~\ref{MD_PE_force}(c)) and $6$~kJ/mol (Fig.~\ref{MD_PE_force}(d)). Here, the potential energy is characterized by significantly larger fluctuations owing to the varying high and low energy phases of square wave potential, resulting in larger forces acting between the two particles compared to the corresponding equilibrium values. The particles are, therefore, subject to frequent and substantial energy changes, driving them to reorient and adjust their positions as they seek stable configurations.
The difference in the force magnitudes between equilibrium and nonequilibrium underscores the mechanism of the increased bond-formation events observed in Fig.~S$7$. These larger forces in the nonequilibrium state catalyze the movement of particles toward each other, expediting the bonding process.

\subsection{Effect of External Drive on Self-Assembly of Large Systems}

In order to demonstrate the efficacy of our proposed design principle of applying an external nonequilibrium drive for a large-scale system, we ran both equilibrium and nonequilibrium simulations of $100$ patchy particles (Section S$7$ in the SI). The particles were modeled with PHS interaction between patches and PBC of the simulation cell to examine how interaction energies $\epsilon_{patch}$ and $\epsilon_{drive}$ influence the structure formation. 
The patches in these simulations were positioned with an interior angle of $135$ degrees for $8$-ring formations and $90$ degrees for $4$-ring formations to explore the ring formation in the resultant assembly structure subjected to different patch orientations, using other parameters similar to those in smaller-scale systems of $8$ patchy particles.

The equilibrium (Fig.~S$8$(a)) and nonequilibrium (Fig.~S$8$(b)) results for the larger system of $100$ particles are consistent with the observations of our smaller systems of $8$, $10$, and $13$ patchy particles, showing faster bond formation in the presence of the drive compared to the corresponding equilibrium results with similar interaction strength between the patches.
Similar observations have been showcased in Figs.~S$9$(a) and S$9$(b) for large systems with $4$-ring specifications.
The overall results are consistent with the small-scale system, demonstrating that using an external drive can facilitate a faster assembly process in the region of weak $\epsilon_{patch}$. Thus, our approach of nonequilibrium drive is independent of the overall system size, effectively negating any finite-size effects and emphasizing the robustness of our assembly protocol.

It is important to note that the formation of rings strictly depends upon the system parameters, particularly on the patch orientation and the number density of the system.
While the system with a patch orientation corresponding to an $8$-ring structure produces open rings or chains  
(Movie~S$11$), the systems with a patch orientation of a $4$-ring produce rings (Fig.~S$9$(c) and Movie~S$12$).
Moreover, adding an external drive to the system in the latter case increases the number of rings formed for weak patchy interactions where no structure formation occurs without the drive (Fig.~S$9$(d)), similar to our previous observations.

We also conducted an equilibrium MD simulation of a $4$-ring specific system comprising $100$ particles, utilizing PHS interactions and PBC for the simulation cell (Fig.~S$10$). This simulation corroborated the structural formations observed in our MC simulation results, as shown in Fig.~S$8$(a), thereby validating the consistency and reliability of our findings across different simulation methodologies.

Since our external drive depends only on the interaction between patches that are nearest neighbors in the predefined target structure, it can only accelerate the structure formation observed in equilibrium simulations, as the nature of the formed structure strictly depends on internal system parameters. Additionally, while increasing the specificity of patchy interactions can impact the optimal interactions for assembly rate and stability timescales, it has no effect on the mechanism of the external drive employed in our design. Therefore, a careful balance of specificity and robustness is necessary for optimal self-assembly processes.

\section{Conclusions}

In our study, we embarked on an exploration to discern the limitations posed by equilibrium conditions in self-assembly processes, and how external drives could alleviate these challenges. Motivated by both biological self-assembly processes (particularly of actin filament and microtubules) and experimental observations, we introduce an external stimulus to a model system of self-assembling patchy particles, to study the resulting dissipative self-assembly. 
To generalize the findings from this study in the future to specific biological systems, patchy particles have been employed here as it has already been shown to closely mimic the dynamic behavior of actin filament and microtubule structure \cite{Ulrich2014, Mahadevan2020}.
Through a synergistic approach combining both MC simulations and MD simulations, under equilibrium and nonequilibrium conditions, our objective was to elucidate the impact of the nonequilibrium external drive on the dynamics of the system. This involved a detailed examination of trajectory realization and an analysis of the fundamental mechanisms driving the observed behaviors.

Our findings reveal that the equilibrium MC simulations of patchy particles with $8$, $10$, and $13$ particles exhibit low assembly rates and low structural stability for weak patchy particle interactions.
The external drive greatly improved the time to assembly and increased the structural stability of the target for the weak interactions regime, reaching the optimal values obtained at equilibrium for a narrow range of the patch interaction energies. This extends the ability of the system to assemble the target for a wider range of conditions compared to the equilibrium scenario, owing to the nonequilibrium external drive \cite{Bisker2018, Ben2021, Faran2023}.
The order parameter ($R$) and total entropy production ($S$) along {with} the realizations demonstrated how the external drive guided the system from a disordered initial state to the order configuration of the target structure.

Our detailed investigation revealed that the complexity of the system influences the self-assembly process in both equilibrium and nonequilibrium conditions. As the complexity increased to a larger number of particles in the system, the effect of the drive was smaller in terms of the final assembly rates and structural stability.
We hypothesize that as complexity increases, the multitude of potential interaction pathways and configurations may lead to kinetic trapping or interference, thereby diminishing the influence of the external stimuli. Consequently, this emphasizes a substantial opportunity for improving the design principles of incorporating external stimuli, ensuring they are dynamically responsive to the increasing complexity of the system.
Furthermore, the impact of patch length and number density on the self-assembly dynamics of patchy particles had no significant effects on the assembly time and structural stability of the target.

To further extend our conclusions, we employed MD simulations of self-assembling patchy particles with a periodic square wave potential as an external drive, inspired by the experimental applications of varying external fields \cite{Song2015,Shields2013}. 
This approach demonstrated similar trends in the acceleration of the assembly and the increase in target stability, mirroring our MC simulation results.
While the effect of the drive brought the system to the optimal performance level at equilibrium, its importance was in the ability to allow for rapid assembly and increased stability in a wider parameter range, i.e., lower values of the patch interaction energy, not possible in equilibrium without the external driving.
Furthermore, the dynamic properties such as translational and rotational diffusivity are influenced by the mass of the particles and patches, but the relative behavior between equilibrium and nonequilibrium conditions remains qualitatively consistent regardless of the mass ratio.
Additionally, an analysis of particle forces and bond dynamics in the presence of the nonequilibrium drive indicated that large momentary forces were responsible for bond formation and stability. Future efforts will include a comprehensive sensitivity analysis of additional system parameters, including the number of particle states, angular dependencies, and varying boundary conditions, to fully understand their impact on self-assembly dynamics.

Our study contributes to the understanding of how external drives can be strategically employed to circumvent the equilibrium limitations in self-assembly processes. 
By employing the LJ potential with different internal states and reflective boundary conditions in MC simulations and the PHS potential with a single internal state and PBC in MD simulations for smaller systems  ($8$, $10$, and $13$ patchy particles), we demonstrate the versatility and effectiveness of our design protocol across varying conditions. This approach highlights the robustness of our protocol in achieving faster and more stable assembly, regardless of the simulation method employed.
Additionally, to ensure that the choice of boundary conditions does not significantly affect our results, we conducted simulations of $100$ particles using PBC in both MC and MD simulations. The outcomes showed similar results, indicating that the qualitative behavior of assembly remains consistent across different boundary conditions.
The use of a square wave as a model for external stimuli opens avenues for experimental implementation, offering insights into experimental observations and applications of patchy particles, which can be implemented across various systems, such as colloids and block copolymers \cite{Gong2017, Li2020, Park2022}.

Looking forward, a deeper understanding of dissipative self-assembly mechanisms will push the development of novel synthetic nanostructures for numerous applications.
Furthermore, the effects of various external forces on the self-assembly process are worth exploring. For instance, alternating electric or magnetic fields could influence particle orientations and interactions, potentially leading to assembly structures overcoming equilibrium limitations. Stochastic perturbation is another choice of external force that might introduce randomness that can hinder or enhance the assembly process. This exploration will also need to consider how these forces are influenced by key parameters such as system size, patch-patch interactions, and binding affinity between different patchy particles to comprehensively understand their impact on self-assembly.
By gaining all these insights, we will extend our approach from simplified models of patchy particles to more complex representations of biological proteins and explore their self-assembly under external influences.

\begin{acknowledgement}

G.B. acknowledges the Zuckerman STEM Leadership Program and the Tel Aviv University Center for AI and Data Science (TAD).
This work was supported by the ERC NanoNonEq $101039127$, the Air Force Office of Scientific Research (AFOSR) under Award No. FA$9550$-$20$-$1$-$0426$, and by the Army Research Office (ARO) under Grant No. W$911$NF-$21$-$1$-$0101$.
The views and conclusions contained in this document are those of the authors and should not be interpreted as representing the official policies, either expressed or 
 implied, of the Army Research Office or the U.S. Government. The authors thank Michael Faran for fruitful discussions.

\end{acknowledgement}

\begin{suppinfo}

The Supporting Information includes detailed sections on the implementation of square wave potential in nonequilibrium MD simulations, comparative dynamics of self-assembly in patchy particle systems, the influence of patch length on self-assembly kinetics and stability, the impact of number density on self-assembly, and the sustained order parameter in nonequilibrium dynamics, featuring:

\textbf{Fig. S$1$}: Order Parameter ($R$) and total Entropy Production ($S$) dynamics for $8$-patchy particles under external drive.
\textbf{Fig. S$2$}: Order Parameter ($R$) and total Entropy Production ($S$) variations in $13$-patchy particles, showing drive and energy effects.
\textbf{Fig. S$3$}: Total Entropy Production ($S$) and order Parameter ($R$) variations in $8$-patchy particles in equilibrium and nonequilibrium for successful realization of the target.
\textbf{Fig. S$4$}: $T_{fas}$ and $T_{stable}$ changes with patch energy and drive in $10$-patch systems under different patch lengths.
\textbf{Fig. S$5$}: Impact of number density on $T_{fas}$ and $T_{stable}$ in $10$-patch systems under different conditions.
\textbf{Fig. S$6$}: Consistent $R$ growth in $8$-patch and $10$-patch systems across square wave potential amplitudes.
\textbf{Fig. S$7$}: Average number of bond formation and dissociation events.
\textbf{Fig. S$8$}:  Variation of $N_{B}$ as a function of MC steps for $100$ particles with $8$-ring specification with and without external drive.
\textbf{Fig. S$9$}:  Variation of $N_{B}$ and $N_{R}$ as a function of MC steps for $100$ particles with $4$-ring specification with and without external drive.
\textbf{Fig. S$10$}:  Variation of $N_{B}$ as a function of MC steps for $100$ particles with $8$-ring specification in equilibrium.

A series of supplementary movies illustrate the dynamic behavior of the systems:
\textbf{Movie S$1$}: Equilibrium MC, $8$ particles, $2$ state.
\textbf{Movie S$2$}: Equilibrium MC, $10$ particles, $2$ states.
\textbf{Movie S$3$}: Equilibrium MC, $13$ particles, $2$ state.
\textbf{Movie S$4$}: Nonequilibrium MC, $8$ particles, $2$ states.
\textbf{Movie S$5$}: Nonequilibrium MC, $10$ particles, $2$ states.
\textbf{Movie S$6$}: Nonequilibrium MC, $13$ particles, $2$ states.
\textbf{Movie S$7$}: Equilibrium MD, $8$ particles, $1$ state.
\textbf{Movie S$8$}: Equilibrium MD, $10$ particles, $1$ states.
\textbf{Movie S$9$}: Nonequilibrium MD, $8$ particles, $1$ state.
\textbf{Movie S$10$}: Nonequilibrium MD, $10$ particles, $1$ states.
\textbf{Movie S$11$}: Equilibrium MC, $100$ particles, $1$ state, $8$-ring specification.
\textbf{Movie S$12$}: Equilibrium MC, $100$ particles, $1$ state, $4$-ring specification.

\end{suppinfo}

\providecommand{\latin}[1]{#1}
\makeatletter
\providecommand{\doi}
  {\begingroup\let\do\@makeother\dospecials
  \catcode`\{=1 \catcode`\}=2 \doi@aux}
\providecommand{\doi@aux}[1]{\endgroup\texttt{#1}}
\makeatother
\providecommand*\mcitethebibliography{\thebibliography}
\csname @ifundefined\endcsname{endmcitethebibliography}  {\let\endmcitethebibliography\endthebibliography}{}

\end{document}


\newpage

\section{Total Entropy Production Calculation in MC Simulations}

The calculation of entropy production in MC simulations involves summing the contributions from each accepted MC move, where each move alters the system's energy. The contribution is calculated by $k_B$ times the log-ratio of the forward and reverse transition probabilities.\cite{Gili_entropy}
Let $P_{forward}$ be the probability of moving forward from one state to another, and $P_{reverse}$ be the probability of moving reverse between the same two states.

 During an MC step, if a particle transitions from an initial energy $( E_{ini} )$ to a final energy $( E_{final} )$, the energy change $( \delta E = E_{final} - E_{ini} )$ determines the transition probabilities. If $\delta E < 0$, indicating a decrease in energy, the forward transition probability, $P_{forward}$, is $1$ and the reverse transition probability, $P_{reverse}$, is $e^{+\delta E / k_B T}$, $T$ is the temperature. Conversely, if $\delta E > 0$, signifying an increase in energy, the forward transition probability, $P_{forward}$, is $e^{-\delta E / k_B T}$, and the reverse transition probability, $P_{reverse}$, is $1$. Thus, the contribution to entropy production, namely $k_B\ln(\frac{P_{forward}}{reverse})$, in the case of $\delta E > 0$ is $-\frac{\delta E}{T}$, and in the opposite case, $\delta E < 0$, it is $+\frac{\delta E}{T}$.
In MC simulations, entropy production accumulates from these probabilistic energy changes and sums to the total entropy change over the course of the simulation. The total entropy production, $S$ is the sum of these contributions. The resulting total entropy production is in the units of the (molar) energy divided by the temperature, kJ / (mol $\cdot$ K).

\section{Self-Assembly Dynamics in $8$ and $13$ patchy Particle Systems in MC Simulations}

We extend the analysis of the order parameter, $R$, and the total entropy production, $S$, to systems of $8$ and $13$ patchy particles, in addition to the $10$ patchy particle system presented in the main text.

\begin{figure}[ht]
    \centering
    \includegraphics[width=16cm]{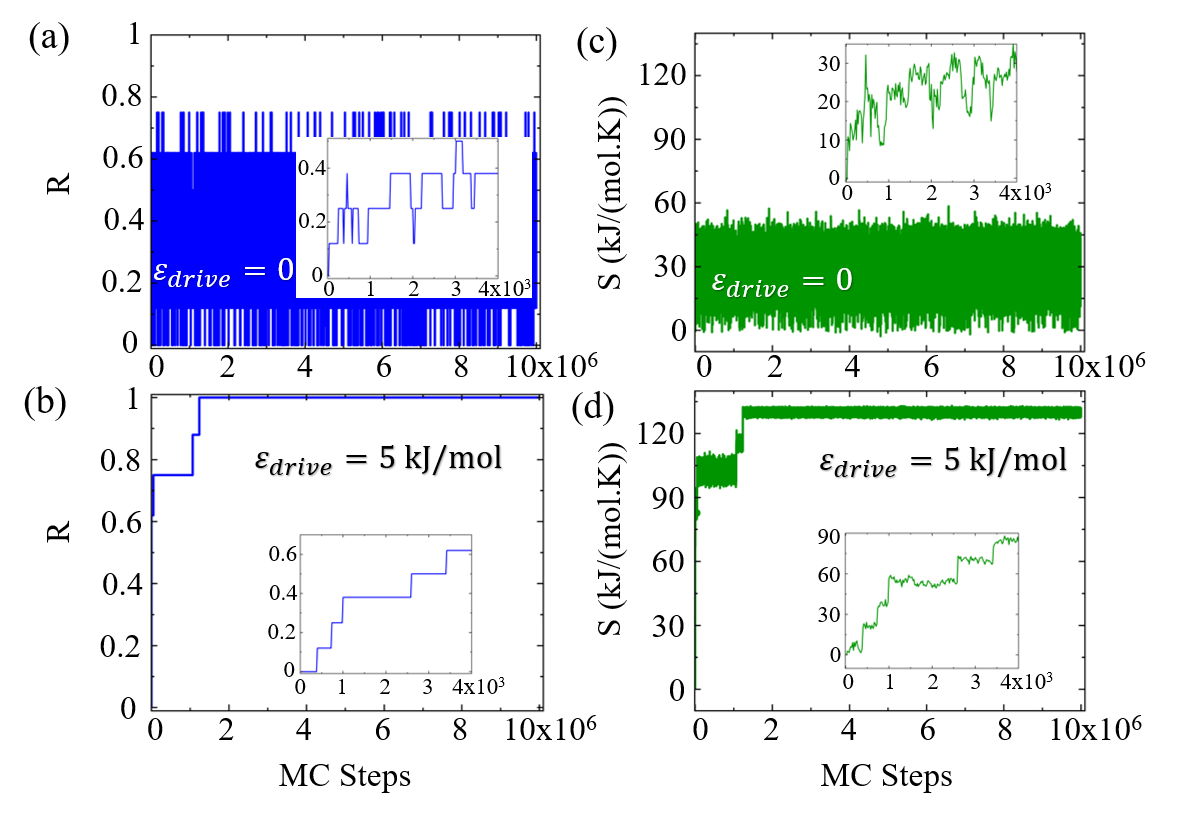}
    \caption{{Results of $8$ patchy particle system.} Order parameter ($R$) as a function of MC steps (a) in equilibrium and (b) under nonequilibrium conditions with an external drive $\epsilon_{drive}=5$~kJ/mol.
    The total entropy production ($S$) as a function of MC steps (c) in equilibrium and (d) under nonequilibrium conditions with an external drive $\epsilon_{drive}=5$~kJ/mol.
    Results are presented for a single MC realization with $\epsilon_{patch}=4$~kJ/mol.
    The inset plots show Zoom in into the early simulation steps.} 
    \label{thermo_prop_8particle}
\end{figure}

The evolution of $R$ for a $8$ patchy particle system under equilibrium ($\epsilon_{drive} = 0$) and nonequilibrium ($\epsilon_{drive} = 5$~kJ/mol) conditions is depicted in Fig.~\ref{thermo_prop_8particle}(a) and Fig.~\ref{thermo_prop_8particle}(b), respectively, for a patchy interaction value $\epsilon_{patch}=4$~kJ/mol. In the absence of an external drive, $R$ displays stochastic behavior, oscillating between $0$ (i.e., no bond formed) and $0.75$ (i.e., $6$ bonds formed out of the $8$ possible bonds), indicating the inability of the system to self-assemble the target structure. 
With the external drive, $R$ displays a rapid transition from a disordered state to the fully assembled target, corresponding to $R=1$ at $1.3\times10^{6}$ MC steps. 
The total entropy production, $S$, without the drive, shows zero entropy production rate (Fig.~\ref{thermo_prop_8particle}(c)), in contrast to the nonequilibrium case where the value of $S$ rapidly increases (Fig.~\ref{thermo_prop_8particle}(d)), reflecting the energy dissipation associated with the use of the external drive.

\begin{figure}[ht]
    \centering
    \includegraphics[width=16cm]{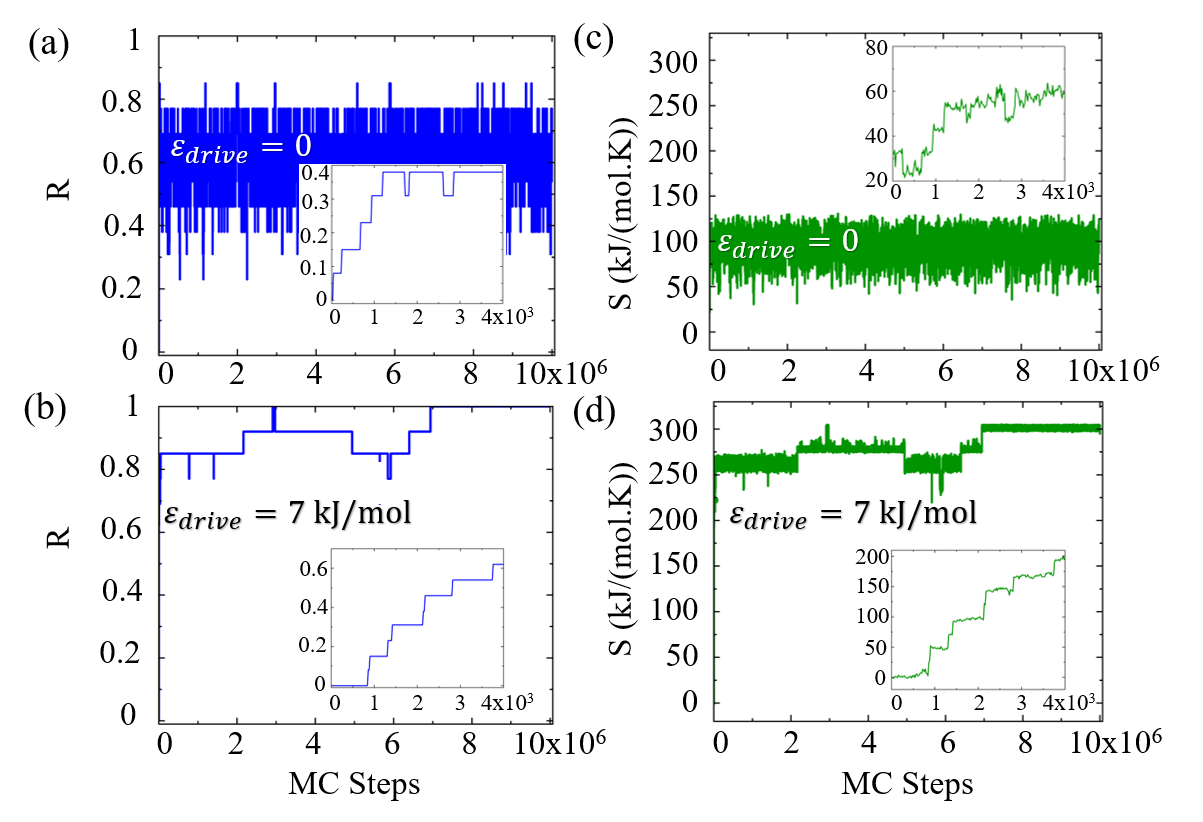}
    \caption{{Results of $13$ patchy particle system.} Order parameter ($R$) as a function of MC steps (a) in equilibrium and (b) under nonequilibrium conditions with an external drive $\epsilon_{drive}=7$~kJ/mol.
    The total entropy production ($S$) as a function of MC steps (c) in equilibrium and (d) under nonequilibrium conditions with an external drive $\epsilon_{drive}=7$~kJ/mol.
    Results are presented for a single MC realization with $\epsilon_{patch}=5$~kJ/mol.}
    \label{thermo_prop_13particle}
\end{figure}

Similarly, for the $13$ patchy particle system with $\epsilon_{patch}=5$~kJ/mol, $R$ fluctuates randomly only between $0.4$ and $0.8$ through the entire course of the simulation under equilibrium conditions (Fig.~\ref{thermo_prop_13particle}(a)), where under nonequilibrium conditions with $\epsilon_{drive}=7$~kJ/mol, $R$ increases in a step-wise manner until it reaches the value of $1$ at $3.9\times10^{6}$ MC steps for the first time, which represents a fully assembled structure (Fig.~\ref{thermo_prop_13particle}(b)). Moreover, $S$ displays fluctuations with mean zero graduate, indicating zero entropy production rate at equilibrium (Fig.~\ref{thermo_prop_13particle}(c)), whereas with the external drive, $S$ rapidly increases (Fig.~\ref{thermo_prop_13particle}(d)).

\begin{figure}[ht]
    \centering
    \includegraphics[width=16cm]{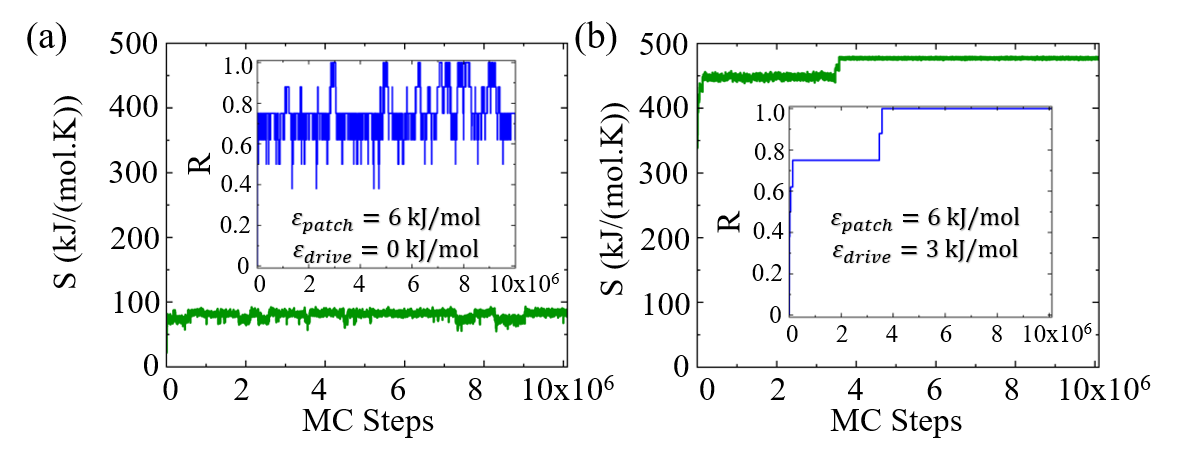}
    \caption{Total entropy production ($S$) as a function of MC steps (a) in equilibrium and (b) under nonequilibrium conditions with an external drive = $3$ kJ/mol. Results are presented for a single MC realization of $8$ particle system with $\epsilon_{patch} = 3$ kJ/mol. The insets show the variation of $R$.}
    \label{equi_nonequi_S}
\end{figure}

These results demonstrate that external driving forces not only facilitate the self-assembly of patchy particles into a predefined structure but also modulate the thermodynamic signatures of the process. This modulation is evident in the transition to an ordered state captured by the increase in $R$ and $S$. 
Furthermore, to provide a comparison, the computation of $S$ and $R$ for both equilibrium and nonequilibrium cases, where each realization reaches the target structure, is plotted in Fig.~\ref{equi_nonequi_S}.

\section{Effect of Patch Length on Self-Assembly Kinetics in MC Simulations}

\begin{figure}[ht]
    \centering
    \includegraphics[width=16cm]{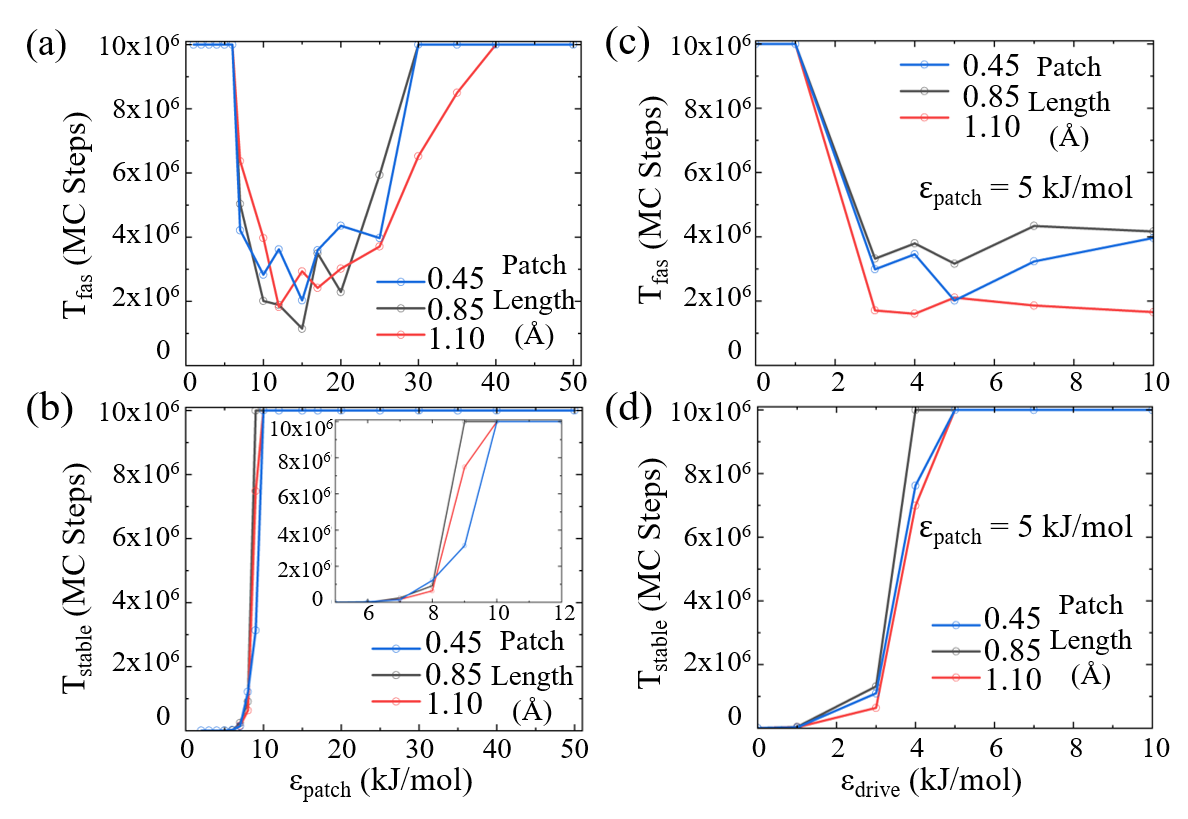}
    \caption{MC simulation results of the $10$ patchy particle system from $20$ independent realizations for patch lengths of $0.4$ \AA\, (blue), $0.85$ \AA\, (black), and $1.1$ \AA\, (red). (a) Median $T_{fas}$ is plotted for different patchy particle interaction energies, $\epsilon_{patch}$, under equilibrium conditions.
    (b) Median $T_{stable}$ is plotted for different patchy particle interaction energies, $\epsilon_{patch}$, under equilibrium conditions. Inset: Zoom into the lower $\epsilon_{patch}$ range. 
    (c) Median $T_{fas}$ is plotted for $\epsilon_{patch}=5$~kJ/mol, as a function of the drive value, $\epsilon_{drive}$, under nonequilibrium conditions.
    (d) Median $T_{stable}$ is plotted for $\epsilon_{patch}=5$~kJ/mol, as a function of the drive value, $\epsilon_{drive}$, under nonequilibrium conditions.}
    \label{patch_length}
\end{figure}

In addition to the the patch length chosen for our main analysis, $0.85$ \AA, we extend our simulations to include patch lengths of $0.4$ \AA\, and $1.1$ \AA\, to investigate the effect of this parameter on the results for the $10$ patchy particle system. 
Under equilibrium conditions, both the median self-assembly time, $T_{fas}$ (Fig.~\ref{patch_length}(a)), and median stability time, $T_{stable}$ (Fig.~\ref{patch_length}(b)), show a similar trend as a function of the patch interaction energy, $\epsilon_{patch}$, for the $3$ patch length values tested.

Under nonequilibrium conditions, with $\epsilon_{patch}=5$~kJ/mol, we also observe a similar trend of $T_{fas}$ (Fig.~\ref{patch_length}(c)) and $T_{stable}$ (Fig.~\ref{patch_length}(d)), showing faster assembly and higher stability, respectively, for increasing drive value $\epsilon_{drive}$, for the various patch lengths.
These results confirm that our conclusions hold irrespective of the chosen patch length parameter value.

\section{Effect of Simulation Cell Dimension on Self-Assembly Kinetics in MC Simulations}

\begin{figure}[ht]
    \centering
    \includegraphics[width=16cm]{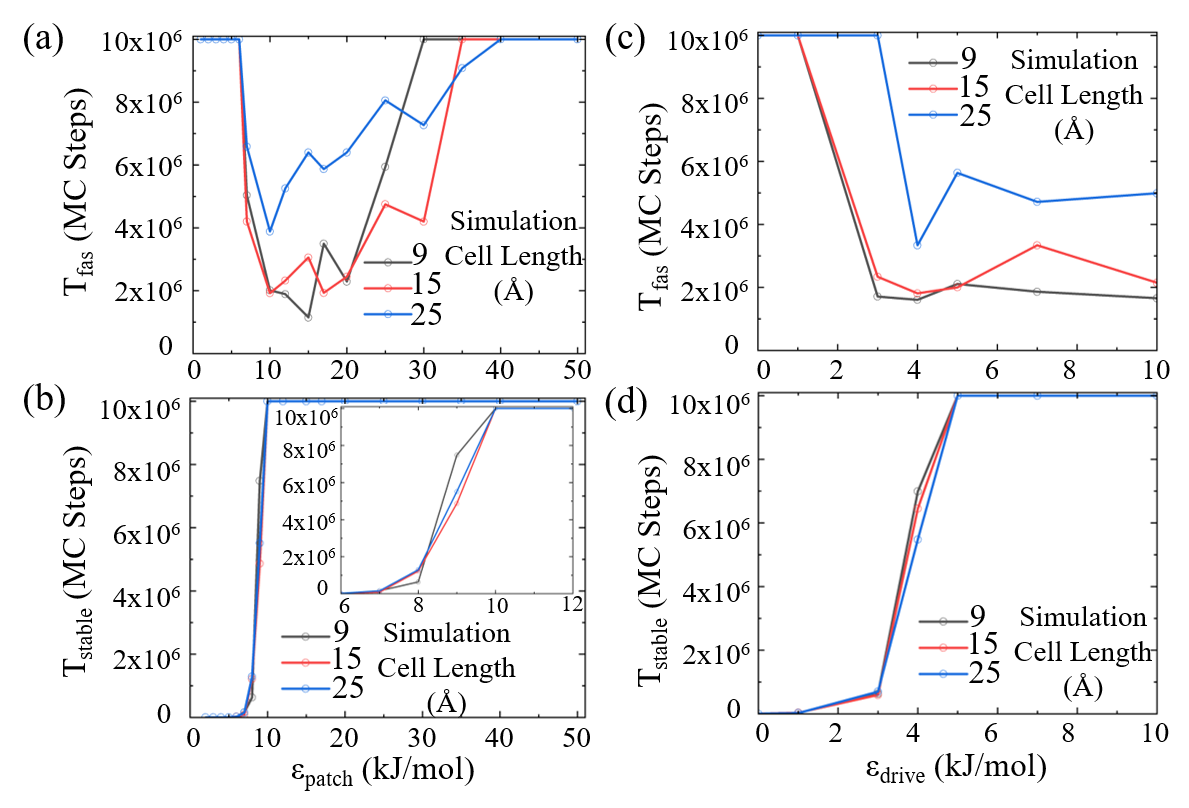}
    \caption{MC simulation results of the $10$ patchy particle system from $20$ independent realizations for different simulation cell dimensions of $9$ \AA\ (black), $15$ \AA\ (red), and $25$ \AA\ (blue). (a) Median $T_{fas}$ is plotted for different patchy particle interaction energies, $\epsilon_{patch}$, under equilibrium conditions.
    (b) Median $T_{stable}$ is plotted for different patchy particle interaction energies, $\epsilon_{patch}$, under equilibrium conditions. Inset: Zoom into the lower $\epsilon_{patch}$ range.
    (c) Median $T_{fas}$ is plotted for $\epsilon_{patch}=5$~kJ/mol, as a function of the drive value, $\epsilon_{drive}$, under nonequilibrium conditions.
    (d) Median $T_{stable}$ is plotted for $\epsilon_{patch}=5$~kJ/mol, as a function of the drive value, $\epsilon_{drive}$, under nonequilibrium conditions.}
    \label{density}
\end{figure}

Understanding the kinetics of self-assembly in the context of number density is important for tuning the properties of assembled structures. In addition to the simulation cell length of $9$~\AA\ used for our main analysis, we include additional simulations for a $15$~\AA\ and $25$~\AA\ cell, for the $10$ patchy particle system (Fig.~\ref{density}). 
Compared to simulation cells of $9$~\AA\ and $15$~\AA, the values of $T_{fas}$ from both equilibrium and nonequilibrium simulations exhibit higher optimal values in the larger system due to the reduced number density, which leads to less frequent interactions among the patchy particles.
Importantly, both $T_{fas}$ and $T_{stable}$ show the same trend in equilibrium and nonequilibrium conditions, further affirming the robustness of our conclusions, which are independent of the simulation cell dimension.

\section{Order Parameter $R$ in Nonequilibrium Molecular Dynamics}

\begin{figure}[!ht]
    \centering
    \includegraphics[width=16cm]{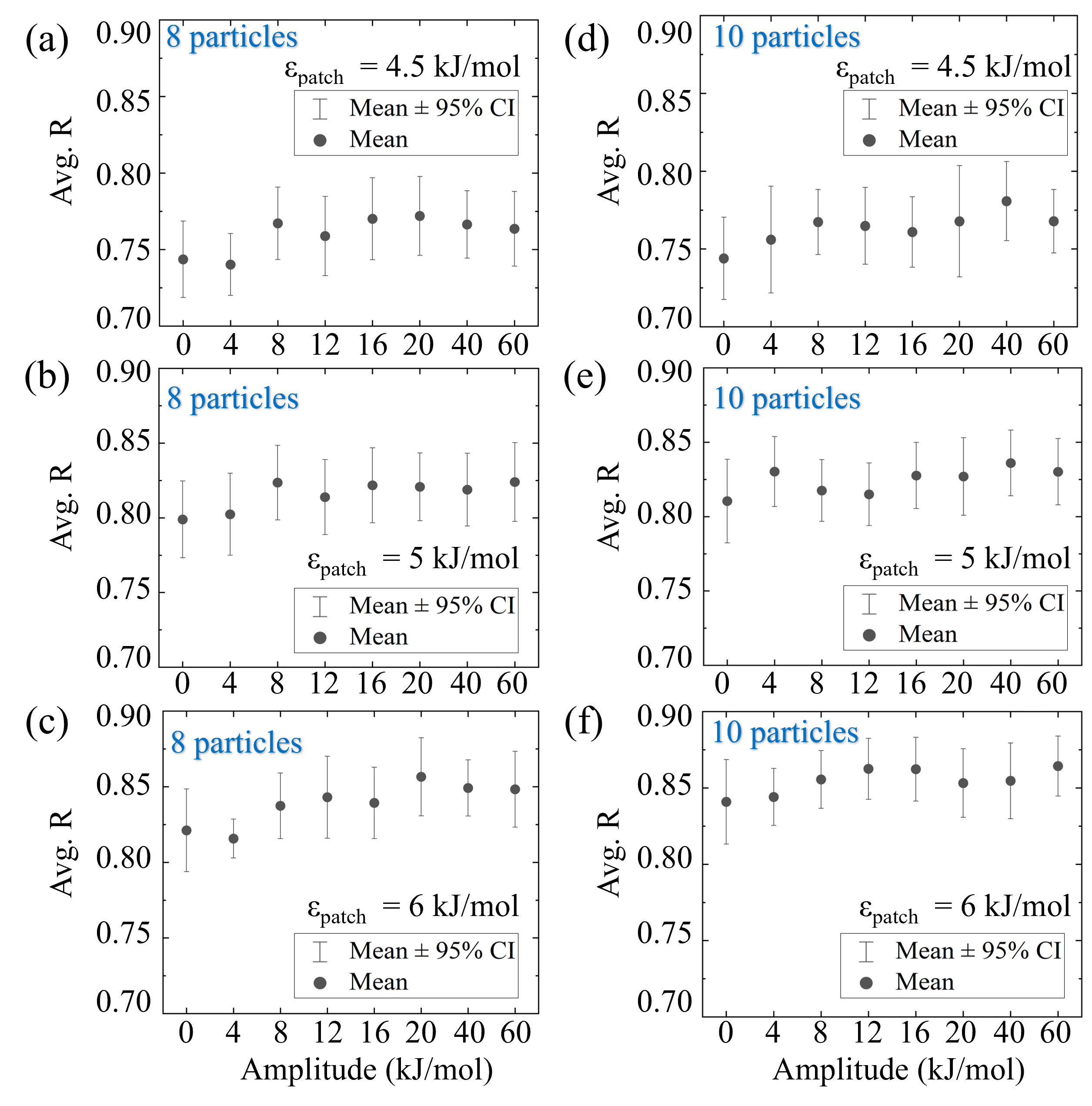}
    \caption{Average order parameter, $R$, from $20$ different MD simulations, as a function of the square wave potential amplitude for $8$ particle system and $\epsilon_{patch}$ value of (a) $4.5$~kJ/mol, (b) $5$~kJ/mol, and (c) $6$~kJ/mol. Average order parameter, $R$, from $20$ different MD simulations, as a function of the square wave potential amplitude for $10$ particle system and $\epsilon_{patch}$ value of (d) $4.5$~kJ/mol, (e) $5$~kJ/mol, and (f) $6$~kJ/mol.} 
    \label{8_10_particle_avg_R}
\end{figure}

The influence of the nonequilibrium driving forces on the self-assembly of patchy particles manifested by the order parameter $R$ is studied for $8$ and $10$ patchy particle systems in equilibrium and nonequilibrium MD simulations. The average order parameter along a realization serves as an indicator of the structural alignment of the system with the desired target assembly. For the $8$ particle (Fig.~\ref{8_10_particle_avg_R}(a)-(c)) and $10$ particle (Fig.~\ref{8_10_particle_avg_R}(d)-(f)) systems, we see a moderate increase in the average $R$ value with increasing amplitude of the square wave potential, for 
$\epsilon_{patch}$ values of $4.5$~kJ/mol, $5$~kJ/mol, and $6$~kJ/mol, mirroring the increase in target stability observed in Figs.~$9$. 

\section{Bond Statistics from MD Simulation}

To investigate the impact of the square wave potential, $U_{square} (t)$, on the assembly kinetics and stability, we track bond formation and dissociation events up to the point of achieving the target structure within our nonequilibrium MD simulations, during both the high energy and low energy phases of the square wave potential.
Note that for a zero amplitude, the system is effectively at equilibrium, and the bond events are averaged over the entire simulation duration and divided by two, for a fair comparison with the nonequilibrium cases in which the events are counted separately for the two potential values.

\begin{figure}[!ht]
    \centering
    \includegraphics[width=16cm]{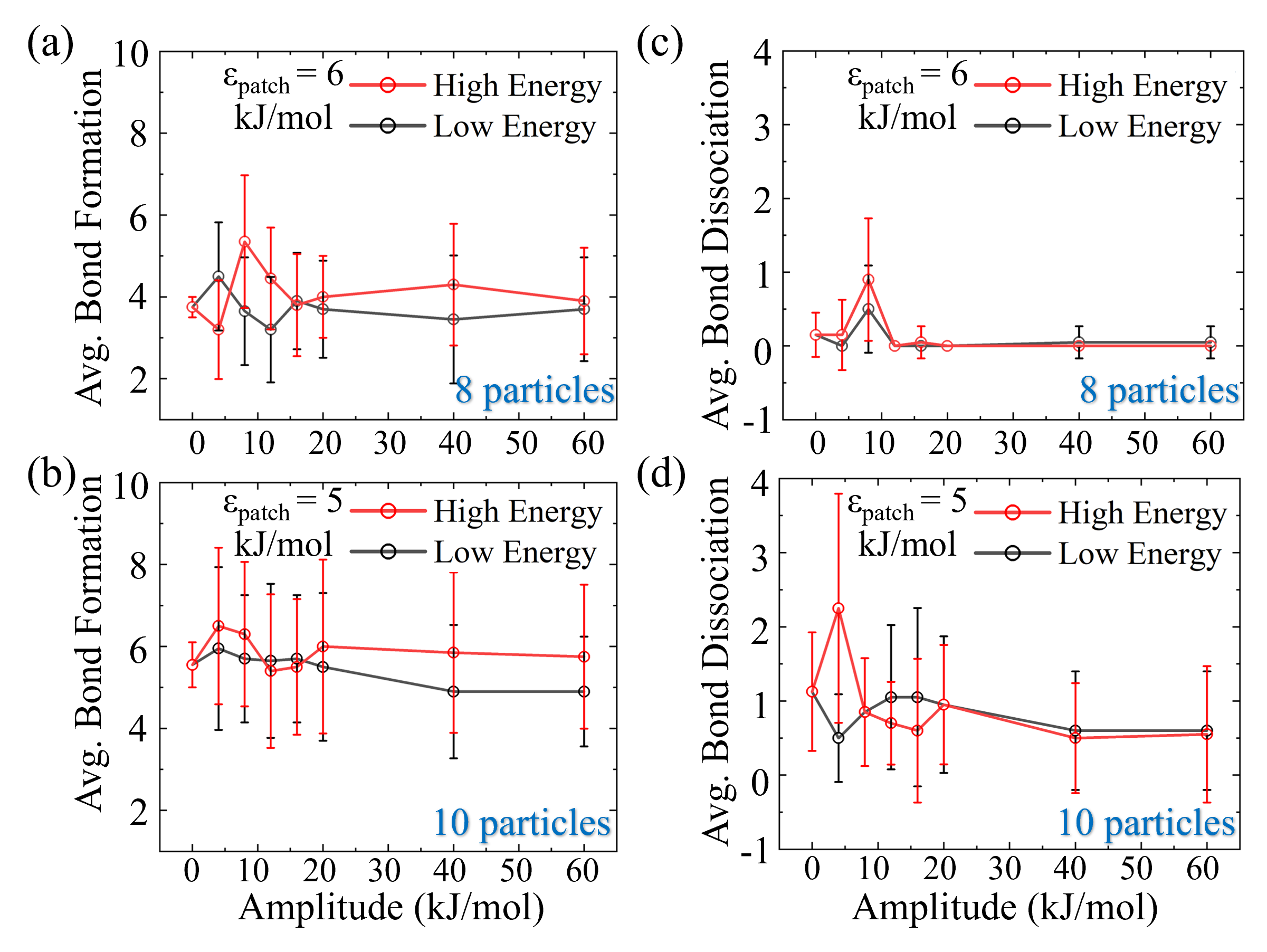}
    \caption{Average number of bond formation events across $20$ individual MD simulations during the high energy (red) and low energy (black) phases of the square wave potential, as a function of its amplitude for (a) $8$ patchy particles and (b) $10$ patchy particles. Similar variations are shown for bond dissociation for (c) $8$ patchy particles and (c) $10$ patchy particles.
    The average number of bond dissociation events across $20$ individual simulations for (a) $8$ patchy particles and (b) $10$ patchy particles. Error bars are standard deviations.}
    \label{8_10_particle_bond}
\end{figure}

{Figs.~\ref{8_10_particle_bond}(a) and \ref{8_10_particle_bond}(b) present the average number of bond formation events for the $8$ particle and $10$ particle systems, for a patchy interaction energy of $6$~kJ/mol and $5$~kJ/mol, respectively. 
At $0$ amplitude, the average number of bonds formed is $\sim 3.8$ for the $8$ patchy particle system and $5.5$ for the $10$ patchy particle system. As the amplitude increases, the number of bond formations tends to oscillate around this equilibrium value with a significant increase in the standard deviation with respect to equilibrium.}

Our analysis indicates that at $0$ amplitude, the average number of bonds formed is  $\sim 3.8$ for the $8$ patchy particle system and $5.5$ for the $10$ patchy particle system. As the amplitude increases, the number of bond formations tends to oscillate around this equilibrium value with a significant increase in the value of standard deviation with respect to equilibrium. Furthermore,  the standard deviation bars for bond formation in the top phase overlap those in the bottom phase, suggesting that while bond formation may appear more frequent in the top phase, the difference is not statistically significant.

{In terms of bond dissociation for the $8$ patchy particles (Fig.~\ref{8_10_particle_bond}(c)), and $10$ patchy particles (Fig.~\ref{8_10_particle_bond}(d)), the average bond dissociation increases from $0.15$ to approximately $0.7$ with increasing drive amplitude from $0$ up to $8$~ kJ/mol for $8$ patchy particles, whereas for the $10$ patchy particle system the average bond dissociation increases from $1.125$ to $2.25$ for the high energy phase for drive amplitude value of $4$~kJ/mol. Beyond these amplitudes, the average bond dissociation reduces to approximately $0$ for the $8$ patchy particles and approximately $0.55$ for the $10$ patchy particle system, and the corresponding standard deviations fall below their equilibrium values.}

{This observation suggests that up to a certain value of drive amplitude, the bond-breaking events increase, whereas, for larger amplitudes, fewer bonds are broken relative to the equilibrium conditions.
Therefore, the overall durability of bonds in our patchy particle system increases in response to the increase in the square wave amplitudes, thereby increasing target stability and facilitating target assembly, as seen in Fig.~$11$ of the main text.} 

\section{Self-Assembly of Large Systems in the Presence of External Drive}

To demonstrate the efficacy of our proposed design principle of employing nonequilibrium force in overcoming equilibrium limitations for larger systems, we present here the equilibrium and nonequilibrium simulations of $100$ patchy particles, with patches modeled to form $8$ and $4$ sided rings. It should be noted that since each particle has two patches and each patch forms one bond, the total number of bonds formed in the system of $100$ particles would be $200$.

\begin{figure}[ht]
    \centering
    \includegraphics[width=16cm]{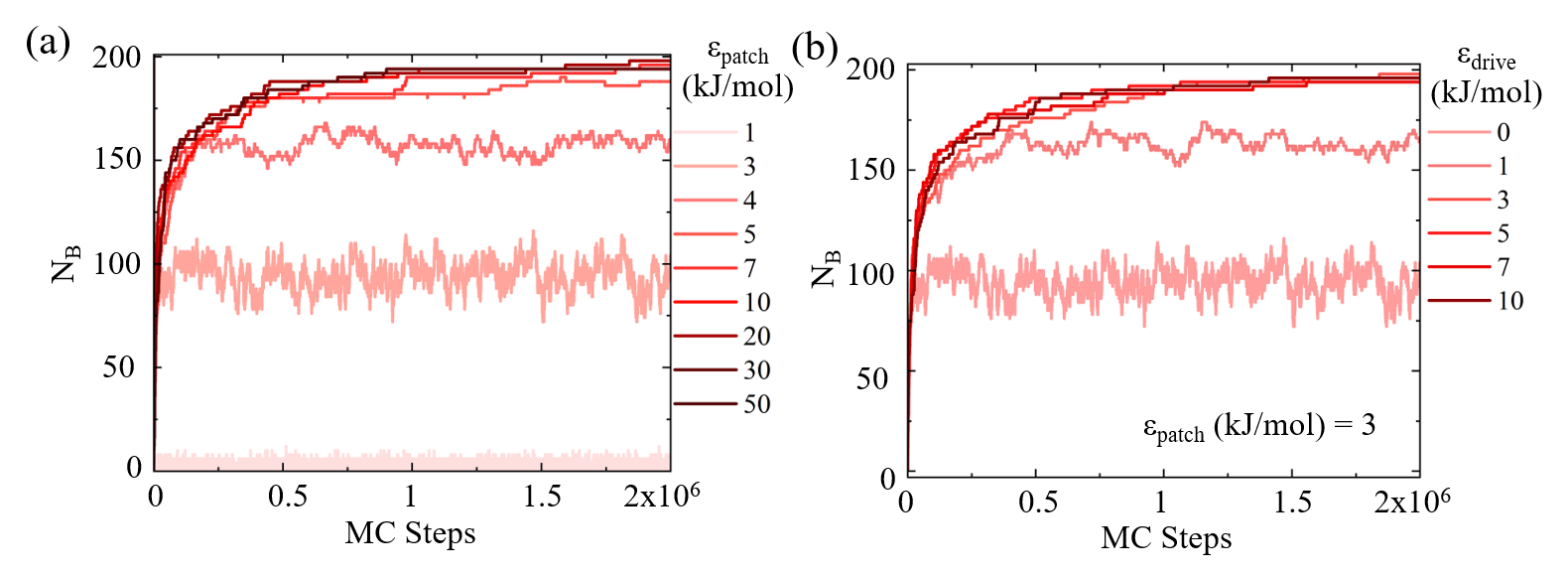}
    \caption{Variation of $N_{B}$ as a function of MC Steps for (a) different values of interaction energy between patches ($\epsilon_{patch}$) in equilibrium and (b) different values of external drive ($\epsilon_{drive}$) for patchy interaction energy of $3$ kJ/mol from single realizations. Two patches are separated from each other by an angular difference of $135^{0}$ (interior angle of the octagon). }
    \label{large_8}
\end{figure}

Fig.~\ref{large_8} illustrates the variation of the number of bonds ($N_{B}$) as a function of MC steps, highlighting the impact of $\epsilon_{patch}$ and $\epsilon_{drive}$. In  Fig.~\ref{large_8}(a), $N_{B}$ is plotted for different values of $\epsilon_{patch}$ in equilibrium. The results show that higher values of $\epsilon_{patch}$ lead to an increased number of bonds, with the number rapidly rising before plateauing at $200$. This indicates that stronger interaction energies promote more stable and numerous bond formations. Fig.~\ref{large_8}(b) presents $N_{B}$ for varying $\epsilon_{drive}$. It shows that applying an external drive increases $N_{B}$ to $200$ while keeping $\epsilon_{patch}$ constant at $3$ kJ/mol. In equilibrium, the number of bonds formed with {the same} $\epsilon_{patch}$ value {(3 kJ/mol)} fluctuates around $\sim 100$. The findings, therefore, reveal that increasing $\epsilon_{drive}$ leads to a higher number of bonds for weak $\epsilon_{patch}$, suggesting that the external drive enhances bond formation beyond equilibrium conditions similar to our earlier observation for small systems. In this system, throughout the equilibrium and nonequilibrium conditions, the resultant structure has the shape of a chain with very infrequent formation of large ring structures of $\sim 20$-sides (see Movie S$11$).

\begin{figure}[ht]
    \centering
    \includegraphics[width=16cm]{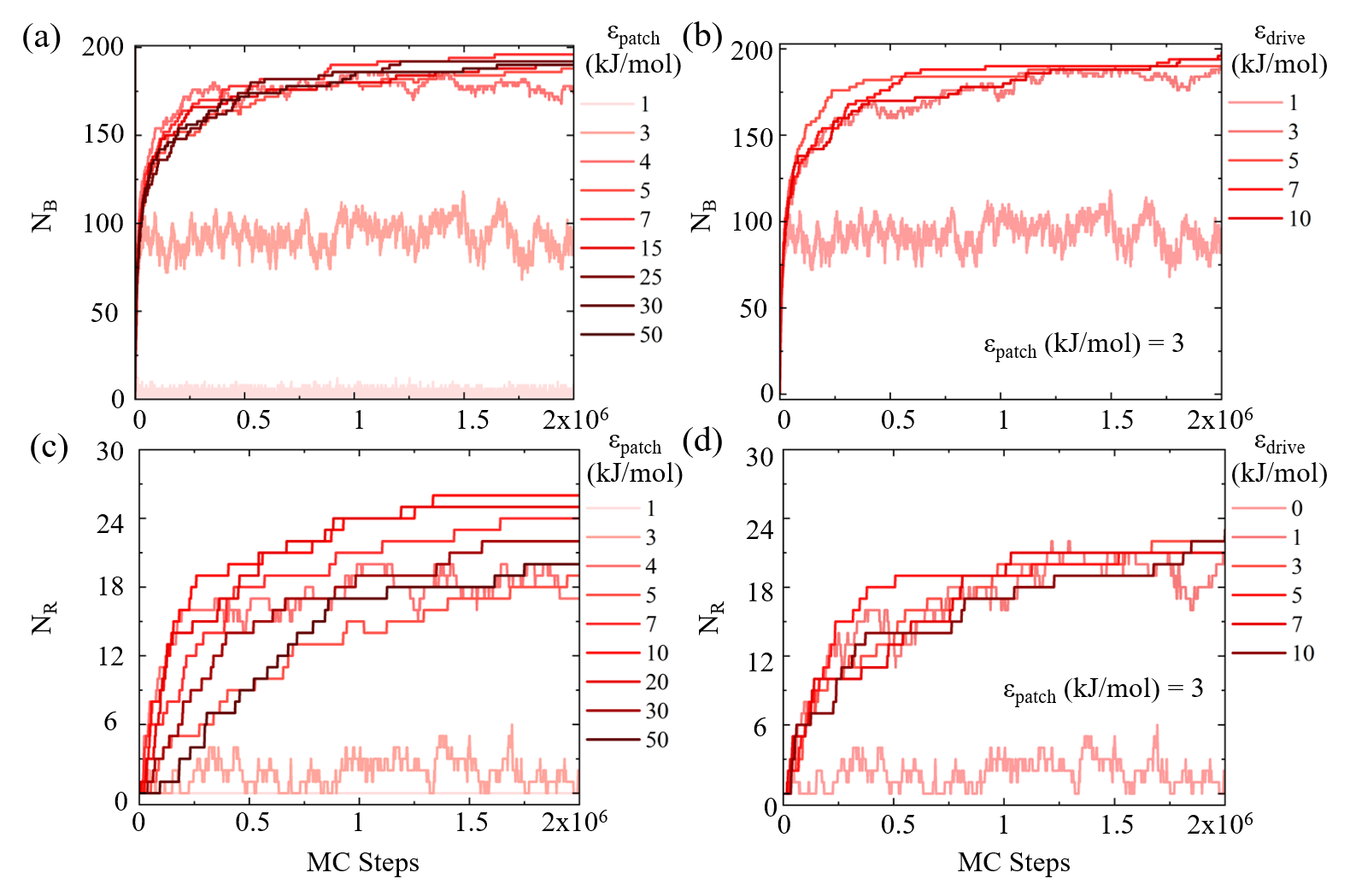}
    \caption{Variation of $N_{B}$ as a function of MC Steps for (a) different values of interaction energy between patches in equilibrium and (b) different values of external drive for patchy interaction energy of $3$ kJ/mol from single realizations. Variation of $N_{R}$ as a function of MC Steps for ({c}) different values of interaction energy between patches in equilibrium and ({d}) different values of external drive for patchy interaction energy of $3$ kJ/mol from single realizations.  Two patches are separated from each other by an angular difference of $90^{0}$ (interior angle of a square).} 
    \label{large_4}
\end{figure}

Fig.~\ref{large_4} illustrates the variation in $N_{B}$ and the number of ring structures ($N_{R}$) as a function of MC steps, with a focus on the impact of the interaction energy between patches and external drives on bond and ring formation. Figs.~\ref{large_4}(a) and \ref{large_4}(b) display $N_{B}$ for different values of interaction energy $\epsilon_{patch}$ in equilibrium and different values of $\epsilon_{drive}$, respectively. In Fig.~$5$, we identified three regions of $T_{fas}$ variation with different $\epsilon_{patch}$ values. In the scenario of a large system, $T_{fas}$ represents the MC step when all the particles form bonds with each other. In Figs.~\ref{large_8}(a) and \ref{large_4}(a), we observe Regions I and II in a larger system: Region I (low $\epsilon_{patch}$, low $N_{B}$) and Region II (increasing $\epsilon_{patch}$, $N_{B}$ approaching $200$). However, Region III is absent in the larger system due to increased particle interactions and mobility, leading to diverse structures like rings and chains and preventing a high-energy overbound state. Thus, the larger system only exhibits Regions I and II, reflecting either no structure formation or stable ring and chain formations.

\begin{figure}[ht]
    \centering
    \includegraphics[width=12cm]{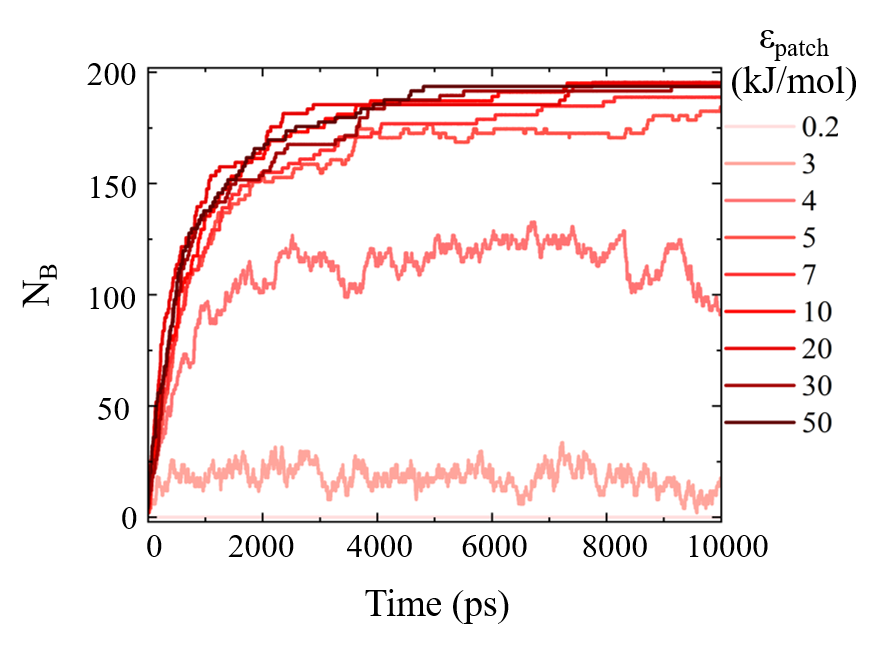}
    \caption{Variation of $N_{B}$ as a function of time for different values of the interaction energy between patches obtained from equilibrium MD simulations. Two patches are separated from each other by an angular difference of $90^{0}$ (interior angle of a square).}
    \label{large_MD}
\end{figure}

The results are similar to what we observed in Fig.~\ref{large_8}. In this system, we observed the formation of a ring structure comprising mainly $4$ and $5$ particles with the infrequent occurrence of a large polygon of $9$ or $10$ sides (see Movie S$12$). Figs.~\ref{large_4}(c) and \ref{large_4}(d) present $N_{R}$, for varying $\epsilon_{patch}$ and then for different $\epsilon_{drive}$ for a specific patch interaction energy of $3$ kJ/mol, respectively. Here, we observe that an increase in values of $\epsilon_{patch}$ resulted in more ring formation (Region II) compared to weak $\epsilon_{patch}$ (Region I), with a significant observation of a lower number of ring formations for higher $\epsilon_{patch}$ value akin to region III of Fig.~$5$ of the main manuscript. Application of $\epsilon_{drive}$ results in a higher number of ring formations compared to equilibrium conditions.

Fig.~\ref{large_MD} illustrates the variation in $N_{B}$ as a function of time for different values of $\epsilon_{patch}$, obtained from equilibrium MD simulations. In this case, the patches are separated by an angular difference of $90$ degrees (interior angle of a square). The plot shows that for lower $\epsilon_{patch}$ values, $N_{B}$ fluctuates around small values indicating that only a few bonds were formed, where more bonds formed over time with increasing $\epsilon_{patch}$ values. Higher values of $\epsilon_{patch}$ lead to a greater maximum $N_{B}$ ($200$ for the system of $100$ particles) and a higher stabilization plateau, indicating more extensive bond formation. Specifically, $\epsilon_{patch} = 50$ kJ/mol results in the highest number of stable bonds, while $\epsilon_{patch} = 0.2$ kJ/mol shows the lowest. This trend is similar to what we observed from Fig.~\ref{large_4}(a) for the same system simulated using MC. This result confirms that the observed qualitative behaviors are consistent across both methods and are intrinsic to the assembly process.

\section{Supplementary movies}

\indent \textbf{Movie S$1$}: Equilibrium MC simulation of $8$ patchy particles,  with $2$ internal states, contained within a cubic box of $8$ \AA\ side length with reflective boundaries. The simulation employs an interaction energy of $3$ kJ/mol for interactions between patches and does not incorporate any external driving forces. Conducted at a temperature of $65$ K, the simulation spans $10\times10^{6}$ MC steps, with the trajectory recorded every $1000^{\text{th}}$ step.

\textbf{Movie S$2$}: Equilibrium MC simulation of $10$ patchy particles, with $2$ internal states, confined within a cubic box of $9$ \AA\ in length and featuring reflective boundaries. The simulation utilizes an interaction energy of $3$ kJ/mol among the patches, with no external driving forces applied. It is conducted at a temperature of $65$ K over $10\times 10^{6}$ MC steps, with the trajectory being captured every $1000^{\text{th}}$ MC step.

\textbf{Movie S$3$}: Equilibrium MC simulation of $13$ patchy particles with $2$ internal states, within a cubic box of $15$ \AA\ in length. The simulation is executed with an interaction energy of $4$ kJ/mol between the patches in the absence of an external driving force. Conducted at $65$ K, the simulation proceeds for $15\times 10^{6}$ MC steps, with trajectory data recorded every $1000^{\text{th}}$ step.

\textbf{Movie S$4$}: Nonequilibrium MC simulation of $8$ patchy particles, with $2$ internal states, contained within a cubic box of 48 \AA\ side length and  reflective boundaries. This simulation is executed with an interaction energy of $3$ kJ/mol between the patches, under the influence of an external driving force of $7$ kJ/mol. Performed at a temperature of $65$ K, the simulation extends over $10\times10^{6}$ MC steps, with the trajectory being recorded every $1000^{\text{th}}$ step.

\textbf{Movie S$5$}: Nonequilibrium MC simulation of $10$ patchy particles, with $2$ internal states, enclosed in a cubic box of $9$ \AA\ side length with reflective boundaries. The simulation utilizes an interaction energy of $3$ kJ/mol between patches, alongside an external driving force of $7$ kJ/mol. Conducted at $65$ K, the simulation proceeds for $10\times 10^{6}$ MC steps, with trajectory data recorded every $1000^{\text{th}}$ step.

\textbf{Movie S$6$}: Nonequilibrium MC simulation of $13$ patchy particles with $2$ internal states, within a cubic box of $15$ \AA\ in length. The simulation is executed with an interaction energy of $4$ kJ/mol between the patches in the presence of an external driving force of $7$ kJ/mol. Conducted at $65$ K, the simulation proceeds for $15\times 10^{6}$ MC steps, with trajectory data recorded every $1000^{\text{th}}$ step.

\textbf{Movie S$7$}: Equlilbrium MD simulation of $8$ patchy particles with $1$ internal state within a cubic box of $8$ \AA\ in length {with periodic boundaries}. The simulation is executed with an interaction energy of $4.5$ kJ/mol between the patches. It is performed at $40$ K for $12000$ ps, with the trajectory being recorded every $2$ ps. 

\textbf{Movie S$8$}: Equlilbrium MD simulation of $10$ patchy particles with $1$ internal state within a cubic box of $9$ \AA\ in length with periodic boundaries. The simulation is executed with an interaction energy of $5$ kJ/mol between the patches. It is performed at $40$ K for $12000$ ps, with the trajectory being recorded every $2$ ps.

\textbf{Movie S$9$}: Nonequlilbrium MD simulation of $8$ patchy particles with $1$ internal state within a cubic box of $8$ \AA\ in length {with periodic boundaries}. The simulation is executed with an interaction energy of $4.5$ kJ/mol between the patches with a square wave potential with an amplitude of $20$ kJ/mol. It is performed at $40$ K for $12000$ ps, with the trajectory being recorded every $2$ ps.

\textbf{Movie S$10$}: This movie demonstrates a nonequlilbrium MD simulation of $10$ patchy particles with $1$ internal state within a cubic box of $9$ \AA\ in length {with periodic boundaries}. The simulation is executed with an interaction energy of $5$ kJ/mol between the patches with a square wave potential with an amplitude of $12$ kJ/mol. It is performed at $40$ K for $12000$ ps, with the trajectory being recorded every $2$ ps.

\textbf{Movie S11}: This movie presents an equilibrium MC simulation of $100$ patchy particles with patches of $8$-ring specification, with $1$ internal state, contained within a cubic box of $30$ \AA\ side length with periodic boundaries. The simulation employs an interaction energy of $20$ kJ/mol for interactions between patches and does not incorporate any external driving forces. Conducted at a temperature of $40$ K, the simulation spans $2 \times 10^{6}$ MC steps, with the trajectory being recorded every $1000^{\text{th}}$ step.

\textbf{Movie S12}: This movie presents an equilibrium MC simulation of $100$ patchy particles with patches of $4$-ring specification, with $1$ internal state, contained within a cubic box of $30$ \AA\ side length with periodic boundaries. The simulation employs an interaction energy of $20$ kJ/mol for interactions between patches and does not incorporate any external driving forces. Conducted at a temperature of $40$ K, the simulation spans $2 \times 10^{6}$ MC steps, with the trajectory being recorded every $1000^{\text{th}}$ step.

\providecommand{\latin}[1]{#1}
\makeatletter
\providecommand{\doi}
  {\begingroup\let\do\@makeother\dospecials
  \catcode`\{=1 \catcode`\}=2 \doi@aux}
\providecommand{\doi@aux}[1]{\endgroup\texttt{#1}}
\makeatother
\providecommand*\mcitethebibliography{\thebibliography}
\csname @ifundefined\endcsname{endmcitethebibliography}  {\let\endmcitethebibliography\endthebibliography}{}